\documentclass[hidelinks,onefignum,onetabnum,nohypdvips]{siamart220329}

\usepackage{jtc-header} 	

\usepackage{lipsum}
\usepackage{amsfonts}
\usepackage{graphicx}
\usepackage{epstopdf}
\usepackage{algorithmic}
\ifpdf
  \DeclareGraphicsExtensions{.eps,.pdf,.png,.jpg}
\else
  \DeclareGraphicsExtensions{.eps}
\fi

\newsiamremark{remark}{Remark}
\newsiamremark{hypothesis}{Hypothesis}
\crefname{hypothesis}{Hypothesis}{Hypotheses}
\newsiamthm{claim}{Claim}

\headers{Linear Discriminant Analysis via Randomized Kaczmarz Method}{J. T. Chi and D. Needell}

\title{Linear Discriminant Analysis with the Randomized Kaczmarz Method\thanks{Submitted to the editors DATE.
\funding{The authors gratefully acknowledge NSF grants DMS-2103093 and DMS-2011140, and the Dunn Family Endowed Fund.}}}

\author{Jocelyn T. Chi\thanks{Department of Applied Mathematics, University of Colorado Boulder, Boulder, CO 
  (\email{jocelyn.chi@colorado.edu}).}
\and Deanna Needell\thanks{Department of Mathematics, University of California at Los Angeles, Los Angeles, CA 
  (\email{deanna@math.ucla.edu}).}
}

\usepackage{amsopn}

\usepackage{graphicx} 
\usepackage{booktabs} 

\ifpdf
\hypersetup{
  pdftitle={Linear Discriminant Analysis with the Randomized Kaczmarz Method},
  pdfauthor={J. T. Chi and D. Needell}
}
\fi

\begin{document}

\maketitle

\begin{abstract}
We present a randomized Kaczmarz method for linear discriminant analysis (rkLDA), an iterative randomized approach to binary-class Gaussian model linear discriminant analysis (LDA) for very large data.  We harness a least squares formulation and mobilize the stochastic gradient descent framework to obtain a randomized classifier with performance that can achieve comparable accuracy to that of full data LDA.  We present analysis for the expected change in the LDA discriminant function if one employs the randomized Kaczmarz solution in lieu of the full-data least squares solution that accounts for both the Gaussian modeling assumptions on the data and algorithmic randomness.  Our analysis shows how the expected change depends on quantities inherent in the data such as the scaled condition number and Frobenius norm of the input data, how well the linear model fits the data, and choices from the randomized algorithm.  Our experiments demonstrate that rkLDA can offer a viable alternative to full data LDA on a range of step-sizes and numbers of iterations.

\end{abstract}

\begin{keywords}
classification, stochastic optimization, supervised learning
\end{keywords}

\begin{MSCcodes}
65F10, 68T10, 68W20, 68W40
\end{MSCcodes}

\section{Introduction}
\label{sec:intro}

Linear discriminant analysis is a classic method of supervised learning, or classification.  Its many applications include classification in genetics \cite{xiong2000computational, witten2011penalized}, facial recognition \cite{liu2002gabor}, pattern recognition \cite{cooke2002two}, hyperspectral imaging \cite{du2007modified}, and chemometrics \cite{guimet2006application}.  Consider a matrix $\mx \in \mathbb{R}^{n \times p}$, where each row contains an observation with measurements on $p$ features.  Each observation $\vx\in \mathbb{R}^{p}$ belongs to one of two classes with class label $y$ so that $y=1$ for Group 1 and $y=0$ for Group 2.  The Gaussian model of linear discriminant analysis (LDA) assumes class-conditional distributions so that $p(\vx \given y)$ is multivariate Gaussian with class means $\vmu_{1} \in \mathbb{R}^{p}$ and $\vmu_{2}$ $\in \mathbb{R}^{p}$ and common covariance $\msig \in \mathbb{R}^{p \times p}$.  Bayes rule then identifies a linear decision boundary for classifying observations to the most likely class.

For moderately sized data in the $n > p$ case, LDA is a classic statistical method of supervised learning.  However, it encounters limitations when either $n$, or both $n$ and $p$, are very large.  If $n$ is large, it may be infeasible to store the entire dataset in working memory on a local machine.  If $p$ is very large, it may be very costly to perform an eigenvalue decomposition on the covariance matrix $\msig$, which is the computational bottleneck in performing LDA. 

To address these computational limitations, we harness the least squares formulation of binary-class LDA and propose LDA with the randomized Kaczmarz method (rkLDA).  The randomized Kaczmarz method (RK) \cite{agaskar2014randomized, censor2009note, dai2013randomized, lin2015learning, liu2016accelerated, liu2014asynchronous, needell2010randomized, needell2016stochastic,  niu2020greedy, nutini2016convergence, steinerberger2021weighted, strohmer2009randomized, zouzias2013randomized} is an instance of stochastic gradient descent (SGD) \cite{robbins1951stochastic} applied to a least squares objective function with a specific step-size that depends on the row chosen in a given iteration \cite{needell2016stochastic}.  By employing a least squares formulation, we can mobilize the well-known SGD \cite{bottou2010large, bottou2012stochastic, nemirovski2009robust} framework to obtain a randomized LDA classifier with comparable performance to that of full data LDA.  Here, ``full data'' refers to the full design matrix $\mx$ and response $\vy$ without randomization or dimension reduction.

Binary classification has numerous important applications in a wide range of domains.  For example, it is widely used to predict outcomes in medicine \cite{Tolles2016}.  While there are numerous examples, we list two to demonstrate its wide applicability.  As an example, it is important to predict whether a patient treated with a stent for coronary artery disease will suffer from in-stent restenosis, or eventual failure of the intervention to reopen an occluded artery \cite{Filatova2022}.  It is also important to predict whether a treated heart failure patient will spend more than a prespecified fraction of follow-up time hospitalized versus at home; this metric has been shown to correlate well with a patient's severity of illness \cite{Desai2020}.  Binary classification problems are also abundant in finance and banking, where it is important to accurately predict loan defaults \cite{fitzpatrick2016empirical, zhu2019study} and bankruptcy \cite{hauser2011predicting, kim2006logistic, laitinen2000bankruptcy} for example.  It also has numerous public policy applications such as in predicting nursing home admissions among the elderly \cite{cai2009factors}, foster care placement \cite{mcdonald2001predicting}, and food assistance program participation \cite{kaiser2008low}.

\subsection{Contributions}

While applying the RK to a least squares formulation of LDA is natural, careful investigation involves several important details.  These include analysis of the expected change in the LDA discriminant function when employing the randomized Kaczmarz solution in lieu of the full data least squares solution that accounts for both algorithmic and model-induced randomness, and investigating the impact of the rkLDA solution on the optimal intercept defined in Section \ref{sec:ls-LDA}.  Our theoretical results in Sections \ref{sec:sketchedLDA} and \ref{sec:proofs} present analysis for the expected deviation of the rkLDA discriminant function on new data from the full data one after a finite number of steps.  Our numerical experiments in Section \ref{sec:experiments} demonstrate that rkLDA can offer a viable alternative to binary-class LDA.

\subsection{Related works}\label{sec:relatedworks}

The least squares formulation of binary-class LDA is well-known (\cite[Section 4.3, Exercise 4.2]{hastie2009elements}, \cite[Section 3.2]{ripley2007pattern}) and has been employed for binary sparse high dimensional discriminant analysis \cite{mai2012direct}.  For more than two classes, a least squares formulation is also available for Fisher's linear discriminant analyis (FDA) \cite{fisher1936use, ye2007least}, which omits distributional assumptions on the data and is not the same as LDA in general.  Other versions of discriminant analysis employing a least squares formulation include optimal scaling \cite{breiman1984nonlinear} and flexible \cite[Section 12.5]{hastie2009elements} discriminant analysis, which enable non-linear transformations of the class labels.

In recent years, \emph{sketching} -- or dimension reduction by random sampling, random projections, or both -- has become popular for large data problems.  Sketching appears to have first appeared as a subspace embedding in \cite{sarlos2006improved}.  Common categorizations of these approaches depend on whether the sketch achieves row compression \cite{boutsidis2009random, ChiIpsenMMLR, ChiIpsenRandLS,  DrineasMM2006, Drineas2011, ipsen2014effect, MMY15, RM16, RT08}, column compression \cite{avron2010blendenpik, kaban2014new, maillard2009compressed, thanei2017random, zhou2007compressed}, or both \cite{LSRN, wang2017sketching}.  

Some works on randomized LDA and its variations exist.  These focus on sketching via subspace embeddings and include feature vector compression \cite{durrant2010compressed}, compression of the observations \cite{lapanowski2021compressing}, fast Johnson-Lindenstrauss transform (FJLT) \cite{ailon2006approximate} projection for Fisher's LDA \cite{ye2017fast}, and random projections in LDA ensembles \cite{durrant2015random}.  These utilize the well-known two-step procedure of first performing efficient dimension reduction with a suitable sketching matrix, and then solving the problem in the reduced dimension.

However, this two-step random projections approach has limited applicability in analyzing massive data.  Although applying a random projection requires less computation time than dense matrix-matrix multiplication when utilizing FJLTs, it still requires $\mathcal{O}(np\log p) + \text{poly}\left(\frac{p}{\epsilon}\right)$ time \cite[page 7]{woodruff2014sketching} when $n > p$ for a desired error $\epsilon$.  Computational time can be drastically reduced if the data are sparse but we consider the general case where the data lack known structure.  Thus, the computation time required can be extraordinarily large for massive data.  Perhaps one version of matrix sketching that could be combined with RK, is randomized block Kaczmarz \cite{needell2014paved, niu2020greedy, rebrova2020block}, where not one row but a block of rows is sampled in each iteration.  We discuss this as potential future work for speeding up convergence in Section \ref{sec:discussion}.

Randomized versions of FDA and its regularized variants also exist \cite{chowdhury2018randomized, tu2014making, ye2017fast}.  In contrast with LDA, FDA does not involve Gaussian assumptions on the data and consequently involves a different objective function.  While the FDA problem is typically solved as a generalized eigenvalue problem, it admits a least squares formulation \cite{ye2007least} that can be employed for multiclass linear discriminant analysis.  Therefore, the RK can also be applied to FDA to obtain a multiclass linear classifier.

Finally, we highlight some other state-of-the-art least squares solvers.  If the design matrix is sparse or has clustered singular values, for example, the conjugate gradient method (CG) is a renowned iterative method that can converge very rapidly \cite{GovL13, hanke1995conjugate, van1986rate}.  We make no such structural assumptions on the data, however, and our numerical experiments show how rkLDA can achieve a sufficiently accurate solution more rapidly than CG on some well-known classification datasets.  Additionally, the RK often outperforms CG \cite[Section 4.2]{strohmer2009randomized} and requires less working memory, making it preferable when memory constraints are an issue.  Another celebrated approach to solving large linear systems is to employ randomized matrix sketching as a preconditioner followed by an iterative least squares solver such as LSQR, as in \cite{avron2010blendenpik}.  While this approach yields fast and highly accurate least squares solutions, one benefit of the RK approach is that the algorithm does not need to observe the entire data.  Therefore, an RK approach opens the doors to future work on a distributed rkLDA to handle data too large to fit in local memory.  We discuss potential adaptations for distributed environments in Section \ref{sec:discussion}.  

\subsection{Notation}
\label{sec:notation}

We assume that vectors are column vectors.   We denote the $j^{th}$ row of a matrix $\mx$ by $\vx_{j}\Tra$, its Frobenius norm by $\|\mx\|_{F}$, and its operator norm by $\|\mx\|_{2}$.  We denote the scaled condition number of $\mx$ by $\kappa(\mx) = \|\mx\|_{F}^{2}\|(\mx\Tra\mx)\Inv\|_{2}$ or $\kappa(\mx) = \|\mx\|_{F}^{2}\|(\mx\Tra\mx)\Dag\|_{2}$ if $\mx$ lacks full column rank.  This is related to the square of the scaled condition number for square matrices in \cite{demmel1988probability}.
Given an integer $n$, denote the set $\{ 1, 2, \dots, n \}$ with $[n]$ and the all ones vector of length $n$ with $\vone_{n}$. 

\section{Gaussian Model Linear Discriminant Analysis (LDA)}
\label{sec:LDA}

We review LDA to underscore the computational difficulties in computing an LDA classifier for very large data with $n > p$.  
Let $\vy \in \mathbb{R}^{n}$ contain class labels for $n$ observations, each belonging to one of $g$ classes.  Let $\vx_{i}$ and $y_{i}$ denote the feature vector and class label for the $i^{th}$ observation.  LDA assumes that observations are class-conditionally Gaussian with class means $\vmu_{k} \in \mathbb{R}^{p}$ for $1 \le k \le g$, common covariance $\msig \in \mathbb{R}^{p \times p}$, and prior probability $\pi_{k}$ of belonging to the $k^{th}$ class with $\sum_{k=1}^{g} \pi_{k} = 1$.  Assume that $\msig$ is nonsingular.  Given a new observation $\vtx \in \mathbb{R}^{p}$, Bayes rule classifies according to
\begin{eqnarray}\label{eqn:gaussianclassification}
	\tilde{y} &=& \arg \max_{k} \delta_{k}(\vtx)
\end{eqnarray}
with the linear discriminant functions (\cite[Section 4.3]{hastie2009elements}, \cite[Section 3.1]{ripley2007pattern})
\begin{eqnarray}\label{eqn:discriminantfuncs}
	\delta_{k}(\vtx) &=& \log \pi_{k} + \vtx\Tra\msig\Inv\vmu_{k} - \frac{1}{2}\vmu_{k}\Tra\msig\Inv\vmu_{k} \quad \text{for} \quad 1 \le k \le g.
\end{eqnarray}

Since $\pi_{k}, \V{\mu}_{k},$ and $\msig$ are unknown, we employ their data-dependent estimates.  Let $n_{k}$ denote the number of observations in the $k^{th}$ class.  Define
\begin{eqnarray*}
	\hat{\pi}_{k} &=& \frac{n_{k}}{n}, \quad 
	\hat{\vmu}_{k} = \sum_{y_{i}=k} \frac{1}{n_{k}}\,\vx_{i}\, , \text{and} \quad
	\hat{\msig} = \frac{1}{n-g}\sum_{k=1}^{g} \sum_{y_{i}=k} (\vx_{i} - \hat{\vmu}_{k})(\vx_{i} - \hat{\vmu}_{k})\Tra.
\end{eqnarray*} 

In practice, we do not compute $\msig\Inv$ explicitly.  Rather than maximizing \eqref{eqn:discriminantfuncs}, we can equivalently minimize
\begin{eqnarray}\label{eqn:discriminantfuncs-v2}
	\delta_{k}^{*}(\vtx) &=& - \delta_{k}(\vtx) + \frac{1}{2}\vtx\Tra \msig\Inv \vtx 
	\,=\, \frac{1}{2}(\vtx - \vmu_{k})\Tra \msig\Inv (\vtx - \vmu_{k}) - \log \pi_{k}
\end{eqnarray}
since the additive term does not depend on $k$.  Let $\msig = \mv\md\mv\Tra$ be an eigenvalue decomposition of $\msig$.  Since $\msig\Inv = \mv\md\Inv\mv\Tra$ exists, we write the first term in \eqref{eqn:discriminantfuncs-v2} as
\begin{eqnarray*}
	\frac{1}{2}(\vtx - \vmu_{k})\Tra \msig\Inv (\vtx - \vmu_{k}) &=& \frac{1}{2}\| \md^{-\frac{1}{2}} \mv\Tra (\vtx - \vmu_{k})\|_{2}^{2} 
	\,=\, \frac{1}{2}\| \vx^{*} - \vmu_{k}^{*}\|_{2}^{2},
\end{eqnarray*}
where $\vx^{*} = \md^{-\frac{1}{2}} \mv\Tra \vtx$ and $\vmu_{k}^{*} = \md^{-\frac{1}{2}} \mv\Tra \vmu_{k}$. 
This enables rewriting \eqref{eqn:discriminantfuncs-v2} as
\begin{eqnarray}\label{eqn:discriminantfuncs-v3}
	\delta_{k}^{*}(\vtx) &=& \frac{1}{2}\| \vtx^{*} - \vmu_{k}^{*}\|_{2}^{2} - \log \pi_{k}.
\end{eqnarray}
Applying $\md^{-\frac{1}{2}} \mv\Tra$ on the left in \eqref{eqn:discriminantfuncs-v3} has a sphering effect since $\md^{-\frac{1}{2}} \mv\Tra \mx$ has identity covariance.  Thus, classifying $\vtx^{*}$ to the group with least $\delta_{k}^{*}$ is equivalent to classifying to the nearest centroid after sphering the data and adjusting for class proportions. 

Therefore, the computational burden in solving \eqref{eqn:gaussianclassification} is in forming $\mhsig$ and computing an eigenvalue decomposition for it (or a singular value decomposition for $\mx$).  This is straightforward for moderately sized data but computationally burdensome if $n$, or both $n$ and $p$, are very large since forming $\mhsig$ requires $\mathcal{O}(np^{2})$ computations and computing its eigenvalue decomposition requires $\mathcal{O}(p^{3})$ computations \cite{pan1999complexity}. 

\subsection{Least Squares Formulation of LDA}\label{sec:ls-LDA}

For binary classification problems, we can employ least squares regression to obtain an LDA classifier in lieu of 
\eqref{eqn:discriminantfuncs-v3}.  Therefore, we focus on this fundamental two-class problem in the $n > p$ case, where \eqref{eqn:gaussianclassification} reduces to the following: we classify $\vtx$ to the second class if and only if
\begin{eqnarray}\label{eqn:binary-decision} 
	\deltalda = \left\{ \vtx - \frac{1}{2}(\vmu_{1} + \vmu_{2}) \right\}\Tra \msig\Inv(\vmu_{2} - \vmu_{1}) + \log\left(\frac{\pi_{2}}{\pi_{1}}\right) > 0.
\end{eqnarray}
Let the $\vbetalda = \msig\Inv(\vmu_{2} - \vmu_{1})$ term in \eqref{eqn:binary-decision} denote the LDA classifier.  Then 
the least squares solution vector is proportional to $\vbetalda$ (\cite[Section 4.3, Exercise 4.2]{hastie2009elements}, \cite[Section 3.2]{ripley2007pattern}).  
Following \cite[Section 4]{fisher1936use}, \cite[Exercise 4.2]{hastie2009elements}, and \cite[Section 3.1]{mai2012direct}, we recode the class labels as $y_{i}=-\frac{n}{n_{1}}$ for observations belonging to Class 1 or $y_{i}=\frac{n}{n_{2}}$ for Class 2, $1 \le i \le n$.  We obtain the Bayes LDA classifier by first solving
\begin{eqnarray}\label{eqn:ls-LDA}
	\begin{pmatrix}\betaho \\ \vhbetals\end{pmatrix} = \arg \min_{\betao, \, \vbeta} \left\{ \frac{1}{2} \sum_{i=1}^{n} (y_{i} - \betao - \vx_{i}\Tra\vbeta)^{2} \right\},
\end{eqnarray}
where $\betao$ is called the \emph{intercept} and $\betaho$  is the least squares estimate for $\betao$.  
Let $\betahopt$ denote an optimal choice for $\betao$.  We classify $\vtx$ to the second class if and only if
\begin{eqnarray}\label{eqn:ls-decisionboundary}
	\deltalslda = \vtx\Tra\vhbetals + \betahopt > 0.
\end{eqnarray}

\subsubsection{Computing the Intercept}\label{sec:intercept}

The LDA classifier in \eqref{eqn:binary-decision} 
does not produce an intercept.  Obtaining a vector $\vhbetals$ that is a scalar multiple of $\vhbetalda$ so that the LDA classification boundary is unchanged, however, requires fitting an intercept in \eqref{eqn:ls-LDA}.  The choice of the intercept affects classification, however, and there are multiple approaches to selecting it.  For example, \cite[Section 4.3]{hastie2009elements} recommends employing a method, such as cross-validation, that minimizes the training error for a given dataset.  Such methods are sometimes referred to as data-driven approaches.  
Meanwhile, \cite[Proposition 2]{mai2012direct} shows that the following closed-form expression for the intercept is optimal in the sense that it minimizes the expected change in the LDA discriminant function for new data
\begin{eqnarray}\label{eqn:optimalbetao}
	\betahopt = -\frac{1}{2}(\vhmu_{1} + \vhmu_{2})\Tra \vhbetals + (\vhbetals)\Tra \mhsig \vhbetals \{(\vhmu_{2} - \vhmu_{1})\Tra \vhbetals \}\Inv \log\left(\frac{n_{2}}{n_{1}}\right).
\end{eqnarray}
We employ the optimal $\betahopt$ from \eqref{eqn:optimalbetao} in our experiments in Section \ref{sec:experiments}.  
Algorithm \ref{alg:lda-ls} summarizes these procedures.  
While the least squares approach to LDA omits Gaussian assumptions \cite{hastie2009elements}, determining an optimal intercept via \eqref{eqn:optimalbetao} assumes class-conditional Gaussian observations.

\begin{algorithm}[ht] 
	\caption{Binary-Class LDA via Least Squares Regression}
	{\bf Input:} Labeled data $(\vy \in \mathbb{R}^{n}, \mx \in \mathbb{R}^{n \times p}$) where the observations belong to one of two classes, and new unlabeled observation $\vtx \in \mathbb{R}^{p}$ \\
	{\bf Output:} Predicted class label $\tilde{y}$ for $\vtx$ 
	\begin{algorithmic}[1]
		\STATE Recode class labels as $y_{i}= \frac{-n}{n_{1}}$ for Class 1 and $y_{i}=\frac{n}{n_{2}}$ for Class 2 for $1 \le i \le n$
		\STATE Compute centroid estimates $\hat{\vmu}_{1}$ and $\hat{\vmu}_{2}$: \\
		$\quad \hat{\vmu}_{k} = \sum_{y_{i}=k} \frac{1}{n_{k}}\vx_{i}$ for $k=1,2$
		\STATE $\mx_{c} = \begin{pmatrix} \vone_{n} & \mx \end{pmatrix}$ \COMMENT{Append $\vone_{n}$ to estimate $\betao$}
		\STATE $\vhbeta = \arg \min_{\vbeta} \frac{1}{2} \| \vy - \mx_{c}\vbeta \|_{2}^{2}$ 
		\STATE Set $\vhbetals$ to be the vector obtained from the last $p$ entries in $\vhbeta$
		\STATE $\betahopt = -\frac{1}{2}(\vmu_{1} + \vmu_{2})\Tra \vhbetals + (\vhbetals)\Tra \mhsig \vhbetals \{(\vmu_{2} - \vmu_{1})\Tra \vhbetals \}\Inv \log\left(\frac{n_{2}}{n_{1}}\right)$
		\STATE Classify $\vtx$ to Class 2 if and only if $\vtx\Tra\vhbetals + \betahopt > 0$
		\RETURN Predicted class label $\tilde{y}$ for $\vtx$
	\end{algorithmic}
	\label{alg:lda-ls}
\end{algorithm}

\section{The Randomized Kaczmarz Method}
\label{sec:randkacz}

For very large data, the RK \cite{agaskar2014randomized,  censor2009note,dai2013randomized, kaczmarz1937angenaherte, lin2015learning, liu2016accelerated, liu2014asynchronous, needell2010randomized, needell2016stochastic, niu2020greedy, nutini2016convergence, steinerberger2021weighted, strohmer2009randomized, zouzias2013randomized} offers a fast, iterative method for approximately solving least squares problems.  
Beginning with an initial estimate $\vbeta_{0}$, the RK proceeds as follows.  In the $k^{th}$ step, we sample the $i^{th}$ observation $i=i_{k}$ i.i.d.\@ at random to select the feature vector $\vx_{i}$ and class label $y_{i}$ according to some sampling distribution $\mathcal{D}$.  We then compute the $(k+1)^{th}$ update as
\begin{eqnarray}\label{eqn:rk}
	\vbeta_{k+1} = \vbeta_{k} + c \cdot \frac{y_{i} - \langle \vx_{i}, \vbeta_{k}\rangle}{\|\vx_{i}\|^{2}_{2}} \, \vx_{i} \, ,
\end{eqnarray}
where $c > 0$ denotes the step-size.  Therefore, the RK is an instance of weighted SGD where the objective function is the least squares objective \cite{needell2016stochastic}. 
If $c=1$ and the rows are sampled with probabilities proportional to their row norms, the RK converges exponentially to a solution that is within a radius of $\vhbetals$ \cite{strohmer2009randomized}.  Other works have shown that appropriate choices of $c<1$ can enable convergence within this radius \cite{censor1983strong, hanke1990acceleration, needell2013two, tanabe1971projection, whitney1967two}.  Finite-iteration guarantees for $n > p$ show a trade-off between the size of the radius and the convergence rate for any choice of $c<1$ in \eqref{eqn:rk} \cite{needell2016stochastic}.

\subsubsection{Importance Sampling} \label{sec:importancesampling}

Selecting rows in \eqref{eqn:rk} with probabilities proportional to their row norms as in \cite{strohmer2009randomized} is an example of \emph{importance sampling}, where greater sampling weights may be placed on some observations.  Importance sampling \cite{needell2016stochastic, nesterov2012efficiency, richtarik2014iteration, strohmer2009randomized} in \eqref{eqn:rk} with any normalized sampling weight $w_{i}\ge 0$ such that $\Exp_{i \sim\mathcal{D}}[w_{i}] = 1$ can be cast in the weighted SGD framework.  If the Lipschitz constants of the SGD gradient estimates have substantially different values, this is preferable to \emph{uniform sampling}, where the rows enjoy equal weights \cite[Section 2]{needell2016stochastic}.

A common variation of importance sampling is \emph{leverage score sampling}, which appears in sketched least squares works \cite{ChiIpsenMMLR, ChiIpsenRandLS, drineas2012fast, ipsen2014effect, MMY15, papailiopoulos2014provable}.  Given a matrix $\mx \in \mathbb{R}^{n \times p}$ with thin singular value decomposition $\mx = \mU\md\mv\Tra$, its row leverage scores are given by the squared Euclidean norms of the rows of $\mU$: 
$\ell_{j} 
= \|\vu_{j}\Tra\|^{2}_{2}$ 
for $1 \le j \le n$.
Leverage scores satisfy $0 \le \ell_{j} \le 1$ and $\sum_{j=1}^{n} \ell_{j} = p$ \cite{drineas2012fast}.

\section{Randomized Kaczmarz LDA}
\label{sec:sketchedLDA}

We harness the least squares formulation of binary-class LDA and propose randomized Kaczmarz LDA (rkLDA).  
Algorithm \ref{alg:lda-rk} presents rkLDA with importance sampling.  Assume i.i.d.\@ weighted random sampling for selecting row $i=i_{k}$ at the $k^{th}$ iteration of \eqref{eqn:rk} with probability $p_{i}$ for $1\le i \le n$ such that $\sum_{i=1}^{n} p_{i} =1$.  
Convergence guarantees for the RK iterates \cite[Corollary 5.1, Corollary 5.2, Corollary 5.3]{needell2016stochastic} hold for the rkLDA iterates $\vbeta_{k}$ conditioned on $\mx$.  

\begin{algorithm}[ht] 
	\caption{Randomized Kaczmarz LDA}
	{\bf Input:} Labeled data $(\vy \in \mathbb{R}^{n}$, $\mx \in \mathbb{R}^{n \times p}$) where the observations belong to one of two classes, new unlabeled observation $\vtx \in \mathbb{R}^{p}$, initial $\vbeta_{0} \in \mathbb{R}^{p+1}$, maximum iterations $K \in \mathbb{N}$, sampling probabilities $p_{i}$ for $1\le i \le n$ with $\sum_{i} p_{i} =1$, and step-size $c < 1$ \\
	{\bf Output:} Predicted class label $\tilde{y}$ for $\vtx$ 
	\begin{algorithmic}[1]
		\STATE Recode class labels as $y_{i}= \frac{-n}{n_{1}}$ for Class 1 and $y_{i}=\frac{n}{n_{2}}$ for Class 2 for $1 \le i \le n$
		\STATE Compute centroid estimates $\hat{\vmu}_{1}$ and $\hat{\vmu}_{2}$: \\
		$\quad \hat{\vmu}_{k} = \sum_{y_{i}=k} \frac{1}{n_{k}}\vx_{i}$ for $k=1,2$
		\STATE $\mx_{c} = \begin{pmatrix} \vone_{n} & \mx \end{pmatrix}$ \COMMENT{Append $\vone_{n}$ to estimate $\betao$}
		\STATE{For $k=0, 1, 2, \dots, K-1$:}
		\STATE $\quad$Randomly sample $i$ from $[n]$ with probability $p_{i}$
		\STATE $\quad \vbeta_{k+1} = \vbeta_{k} + c \cdot \frac{y_{i} - \langle \vx_{c_{i}}, \vbeta_{k}\rangle}{\|\vx_{c_{i}}\|^{2}_{2}} \, \vx_{c_{i}}$
		\STATE Set $\vhbetark$ to be the vector obtained from the last $p$ entries of $\vbeta_{K}$
		\STATE Select $\betahopt$ via \eqref{eqn:optimalbetao}
		\STATE Classify $\vtx$ to Class 2 if and only if $\vtx\Tra\vhbetark + \betahopt > 0$
		\RETURN Predicted class label $\tilde{y}$ for $\vtx$
	\end{algorithmic}
	\label{alg:lda-rk}
\end{algorithm}

Each iteration in Algorithm \ref{alg:lda-rk} requires $\mathcal{O}(p)$ flops.  Runtime experiments in Section \ref{sec:exp2} give empirical evidence of the number of iterations required in the context of classification.

\subsection{Quantifying the expected difference between the rkLDA and full data LDA discriminant functions on new data}\label{sec:lda-theory}

Consider new unlabeled data $\mtx \in \mathbb{R}^{N \times p}$.  
A primary question is: How does the LDA discriminant function on new data $\mtx$ change if we employ the RK solution $\vhbetark$ from Algorithm \ref{alg:lda-rk} in place of the full data solution $\vhbetals$ from Algorithm \ref{alg:lda-ls}?  To answer this question, we consider three sources of randomness: 1) \emph{algorithmic randomness} from the RK iterates $\{ i\}$, which are sampled according to some sampling distribution $\mathcal{D}$; 2) \emph{model-induced randomness of the training data} -- the class-conditional distribution of the training data $\mx$; and 3) \emph{model-induced randomness of the new data} -- the class-conditional distribution of the new unlabeled data $\mtx$.  Therefore, we focus on the expected change in the LDA discriminant function conditioned on the training data at the $k^{th}$ step of Algorithm \ref{alg:lda-rk}:   
$\Exp \left[\|\mtx\vhbeta_{k} - \mtx\vhbetals \|^{2}_{2} \given \mx \right]$.
The conditional expectation is with respect to both algorithmic randomness from RK and model-induced randomness of the new data.  

Theorem \ref{thm:expecteddiffXbeta} presents the expected change from the last $p$ entries in $\mtx\vhbeta_{k}$ and $\mtx\vhbetals$ for any sampling probability $p_{i}$ such that ${\sum_{i=1}^{n} p_{i} =1}$.  These are estimated with the intercept term included in the regression model but do not include the optimal intercept since it is typically computed via a separate post-processing as described in Section \ref{sec:intercept}.  Section \ref{sec:diffbetaopt} discusses the expected change in $\betahopt$ from \eqref{eqn:optimalbetao} if estimated with $\vhbetark$ rather than $\vhbetals$.  
For notational simplicity, let the first column of the training data $\mx \in \mathbb{R}^{n \times (p+1)}$ be $\vone_{n}$.

\begin{theorem}\label{thm:expecteddiffXbeta}
	Given training data $\mx \in \mathbb{R}^{n \times (p+1)}$, where the first column of $\mx$ is $\vone_{n}$, and recoded class labels $\vy \in \mathbb{R}^{n}$ as in Algorithm \ref{alg:lda-rk}, let $\vhbeta$ be the minimizer of
	$
	\vhbeta = \arg \min_{\vbeta} \frac{1}{2} \| \mx \vbeta - \vy\|^{2}_{2},
	$
	and let $\vhbeta_{k}$ and $\vhbetals$ be the vectors formed from the last $p$ entries in $\vbeta_{k}$ from Algorithm \ref{alg:lda-rk} and from $\vhbeta$, respectively.  Let $p_{i}$ for $1\le i \le n$ with $\sum_{i} p_{i} =1$ denote sampling probabilities in Algorithm \ref{alg:lda-rk}.  
	For notational simplicity, let $\alpha_{i} = \frac{\|\vx_{i}\|^{2}_{2}}{p_{i}}$ for $1\le i \le n$ and let $\tilde{\alpha}$ be a lower bound for $\alpha_{i}$ so that $\tilde{\alpha} \le \alpha_{i} \le \sup_{i} \alpha_{i} = \|\mx\|^{2}_{F}$ almost surely.
	For any new unlabeled data $\mtx \in \mathbb{R}^{N \times p}$, $\mtx \vhbeta_{k}$ from Algorithm \ref{alg:lda-rk}, where the $i^{th}$ observation is selected at the $k^{th}$ step with probability $p_{i}$ satisfies
	\begin{eqnarray}\label{eqn:mainresult}
		\Exp\left[\|\mtx\vhbeta_{k} - \mtx\vhbetals \|^{2}_{2} \given \mx\right]
		&\le& \left( 1 - 2\frac{c }{\kappa(\mx)} \Big(1-\frac{c}{\tilde{\alpha}} \|\mx\|^{2}_{F}\Big) \right)^{k} \Exp\|\mtx\|_{2}^{2}\; \|\vhbeta_{0} - \vhbetals\|^{2}_{2} \nonumber\\
		&&+ \frac{c}{\tilde{\alpha}}\, \frac{\kappa(\mx)}{(1-\frac{c}{\tilde{\alpha}} \|\mx\|^{2}_{F})} \Exp\|\mtx\|_{2}^{2}\,r^{\star} ,
	\end{eqnarray}
	where $c < \frac{\alpha_{i}}{\sup_{i} \alpha_{i}} < 1$ is a fixed step-size, and $r^{\star} = \|\mx\vhbeta - \vy\|^{2}_{2}$.
	The conditional expectation is with respect to the sampling distribution of the iterates $\{i\}$ and the class-conditional distribution of $\mtx$.
\end{theorem}


Theorem \ref{eqn:mainresult} shows the expected change in the LDA discriminant function if we employ the RK solution in lieu of the full-data least squares solution.  In particular, Theorem \ref{eqn:mainresult} shows how this expected difference depends on the following: 1) quantities that depend on the data such as the conditioning of $\mx$, 2) how well the linear model fits the data (as quantified by $r^\star$), and 3) choices in the RK algorithm such as the step-size $c$, the sampling probabilities $p_{i}$, and the number of iterates $K$.

The last term in \eqref{eqn:mainresult} contains the radius, or convergence horizon, described in Section \ref{sec:intro}.  This horizon is positive given the definitions of $\tilde{\alpha}$ and $c$.  Additionally, the dependence on $c$ enables a trade-off between the size of the horizon and convergence speed: a smaller $c$ enables a smaller horizon but also results in slower convergence.  This radius depends on elements that are independent of the algorithm for any given dataset with labels $\vy$, training data $\mx$, and new unlabeled data $\mtx$.  For example, the size of the radius depends on how close $\vy$ is to the column space of $\mx$ as indicated by the residual $r^{\star}$ and the conditioning of $\mx$.  
In particular, we observe how Theorem \ref{thm:expecteddiffXbeta} depends on the conditioning of $\mx$.  Notice that from \cite[(3)]{strohmer2009randomized} we obtain $1 \le \frac{\kappa(\mx)}{p+1} \le k^{2}(\mx)$, where $k(\mx)$ is the traditional condition number defined by the ratio of the largest singular value of $\mx$ to its smallest singular value.  If $\mx$ is perfectly conditioned so that $k(\mx) = 1$, then $\kappa(\mx) = p+1$.  Moreover, $\kappa(\mx)$ increases as conditioning worsens.  Therefore, the bound in Theorem \ref{thm:expecteddiffXbeta} is larger when $\mx$ is poorly conditioned and is smaller when $\mx$ is well-conditioned.

We next observe the number of iterations $k$ required to achieve an expected discriminant function error tolerance $\epsilon$.  This follows directly from Theorem \ref{thm:expecteddiffXbeta} and holds for any desired tolerance $\epsilon > 0$ and suitably chosen sampling probabilities $p_{i}$.

\begin{corollary}\label{cor:cor-iterationsk}
	Given a tolerance $0 < \epsilon < \frac{1}{2}$, a step-size of 
	\begin{eqnarray*}
		c = \frac{\epsilon \, \tilde{\alpha}}{2\, \kappa(\mx) \Exp\|\mtx\|^{2}_{2}\,+ 2\, \epsilon \|\mx\|^{2}_{F}}
	\end{eqnarray*}
	is sufficient to ensure that after
	\begin{eqnarray}\label{eqn:k-iterations}
		k \ge \log\Big( \frac{2 \epsilon_{0} \Exp\|\mtx\|^{2}_{2} }{\epsilon} \Big) 
		\left( \frac{\|\mx\|^{2}_{F}}{4\Exp\|\mtx\|^{2}_{2}} + \frac{\kappa(\mx)}{2\epsilon} \right)
	\end{eqnarray}
	iterations in Algorithm \ref{alg:lda-rk}, we have \mbox{$\Exp \Big[\|\mtx\vhbeta_{k} - \mtx\vhbetals\|^{2}_{2} \given \mx\Big] < \epsilon (1 + r^\star)$}, where 
	$\epsilon_{0} = \|\vhbeta_{0} - \vhbetals\|^{2}_{2}$, $r^{\star} = \|\mx\vhbeta - \vy\|^{2}_{2}$,
	and the expectation is over the random sampling in Algorithm \ref{alg:lda-rk} as well as the randomness in $\mtx$.
\end{corollary}

Compared with \cite[Theorem 2.1]{needell2016stochastic}, which holds generally for any $\mu$-strongly convex function, Theorem \ref{thm:expecteddiffXbeta} focuses on the least squares formulation for binary-class LDA in \eqref{eqn:ls-LDA}.  Since LDA is a classification problem, we focus on $\mtx\vhbeta_{k}$ on new unlabeled data at the $k^{th}$ iterate rather than the $k^{th}$ iterate $\vhbeta_{k}$.    
Additionally, Theorem \ref{thm:expecteddiffXbeta} accounts for the Gaussian modeling assumptions on the training data $\mx$ and unlabeled data $\mtx$.  Therefore, the conditional expectation in Theorem \ref{thm:expecteddiffXbeta} is with respect to the model-induced randomness of $\mtx$ after accounting for the algorithm-induced randomness from the iterates in \eqref{eqn:rk}. 
Since we account for randomness in $\mtx$, the first term in Theorem \ref{thm:expecteddiffXbeta} depends additionally on $\Exp\|\mtx\|^{2}_{2}$.  

Since many results in this setting are ``worst-case" results and often overly pessimistic, in theory, a large $k$ might be required to guarantee a very small expected discriminant function error.  In practice, however, our numerical experiments demonstrate that very good classification results can be obtained even when $k$ is quite small.  In our real data experiments with larger $p$ in Section \ref{sec:exp2}, for example, fewer than $2{,}500$ iterations were required to obtain performance that was very comparable to LDA.

We make some observations about $\Exp \| \mtx\|_{2}$, the expected largest singular value of new data $\mtx$.  For large $N$ and $p$ and standardized data such that the entries of $\mtx$ are i.i.d. $\mathcal{N}(0,1)$, one can employ asymptotic results on the largest singular value of $\mtx$ from \cite[Proposition 6.1]{edelmanthesis}. Namely, if the entries of $\mtx$ are i.i.d. $\mathcal{N}(0,1)$, and $N$ and $p$ tend to infinity such that $\frac{N}{p}$ tends to a limit $d \in [0,1]$, then $\frac{1}{\sqrt{N}}\|\mtx\|_{2}$ converges to $1 + \sqrt{d}$ almost surely.  It follows that if we additionally have that $N=p$ so that $d=1$, then $\|\mtx\|_{2}$ converges to $2\sqrt{N}$ almost surely.  Additionally, if $N=p$ and the entries of $\mtx$ are i.i.d. $\mathcal{N}\left(0, \frac{1}{N}\right)$, then \cite[Theorem II.4]{davidson2001local} offers the following bounds on $\|\mtx\|_{2}$
\begin{eqnarray*}
	\mathcal{P}\left( \|\mtx\|_{2} \ge s_{u} \right) \le \exp(-\gamma s_{u}^{2}N^{2})
	\quad \text{and} \quad
	\mathcal{P}\left( \|\mtx\|_{2} \le s_{l} \right) \le (C s_{l})^{N^{2}},
\end{eqnarray*}
where $s_{u} \le s_{0}, s_{l} \ge 0$, and $\gamma, C$, and $s_{0}$ are universal positive constants.  These bounds provide a streamlined view of how the largest singular value of the new data $\mtx$ depends on $N$ under some simplified conditions.

The following corollaries illustrate Theorem \ref{thm:expecteddiffXbeta} for some common choices of sampling probabilities $p_{i}$.  Table \ref{table:results_with_sampling} at the end of this section summarizes their differences.   Corollary \ref{cor:cor1} holds for \emph{row weight sampling}, where $p_{i} = \frac{\|\vx_{i}\|^{2}_{2}}{\|\mx\|^{2}_{F}}$ (\cite[Corollary 5.1]{needell2016stochastic} and \cite[Section 2]{strohmer2009randomized}).

\begin{corollary}\label{cor:cor1}
	Given training data $\mx \in \mathbb{R}^{n \times (p+1)}$, where the first column of $\mx$ is $\vone_{n}$, and recoded class labels $\vy \in \mathbb{R}^{n}$ as in Algorithm \ref{alg:lda-rk}, let $\vhbeta$ be the minimizer of
	$
	\vhbeta = \arg \min_{\vbeta} \frac{1}{2} \| \mx \vbeta - \vy\|^{2}_{2},
	$
	and let $\vhbeta_{k}$ and $\vhbetals$ be the vectors formed from the last $p$ entries in $\vbeta_{k}$ from Algorithm \ref{alg:lda-rk} and from $\vhbeta$, respectively.
	Let $c<1$.  For any new unlabeled data $\mtx \in \mathbb{R}^{N \times p}$, $\mtx \vhbeta_{k}$ from Algorithm \ref{alg:lda-rk}, where the $i^{th}$ observation is selected at the $k^{th}$ step with probability $p_{i}=\frac{\|\vx_{i}\|^{2}_{2}}{\|\mx\|^{2}_{F}}$ satisfies
	\begin{eqnarray}\label{eqn:cor1b}
		\Exp\left[\|\mtx\vhbeta_{k} - \mtx\vhbetals \|^{2}_{2} \given \mx\right]
		&\le& \left( 1 - 2c\frac{(1-c)}{\kappa(\mx)} \right)^{k} \Exp\|\mtx\|_{2}^{2} \, \|\vhbeta_{0} - \vhbetals\|^{2}_{2} \\
		&+& \frac{c}{1-c} \, \kappa(\mx) \Exp\|\mtx\|_{2}^{2} \, r, \nonumber
	\end{eqnarray}
	where $r = \frac{\|\mx\vhbeta - \vy\|^{2}_{2}}{\|\mx\|^{2}_{F}}$.
	The conditional expectation is with respect to the sampling distribution of the iterates $\{i\}$ and the class-conditional distribution of $\mtx$.
\end{corollary}

Similar to \cite[Corollary 5.1]{needell2016stochastic} and \cite[Theorem 2]{strohmer2009randomized}, Corollary \ref{cor:cor1} does not depend on the number of observations.  Rather, convergence with row weight sampling depends on the conditioning of $\mx$, the distance between the initial estimate $\vhbeta_{0}$ and the minimum $\vhbetals$, the least squares residuals at the minimum in $r$, and $\Exp\|\mtx\|^{2}_{2}$.   
Notice that if $\vy$ is close to the column space of $\mx$ so that the least squares residuals in $r$ are small, then the convergence horizon in \eqref{eqn:cor1b} is likewise smaller.
For very large $n$, pre-computing the row norms may require batched computations on subsets of the rows of the training data.  For comparison, Corollary \ref{cor:cor2} illustrates Theorem \ref{thm:expecteddiffXbeta} with \emph{uniform sampling}, where $p_{i} = \frac{1}{n}$ so that all rows are selected with equal probabilities.

\begin{corollary}\label{cor:cor2}
	Given training data $\mx \in \mathbb{R}^{n \times (p+1)}$, where the first column of $\mx$ is $\vone_{n}$, and recoded class labels $\vy \in \mathbb{R}^{n}$ as in Algorithm \ref{alg:lda-rk}, let $\vhbeta$ be the minimizer of
	$
	\vhbeta = \arg \min_{\vbeta} \frac{1}{2} \| \mx \vbeta - \vy\|^{2}_{2},
	$
	and let $\vhbeta_{k}$ and $\vhbetals$ be the vectors formed from the last $p$ entries in $\vbeta_{k}$ from Algorithm \ref{alg:lda-rk} and from $\vhbeta$, respectively.
	Let  $c < \min\Big(1, \frac{n}{\kappa(X)}\Big)$.  For any new unlabeled data $\mtx \in \mathbb{R}^{N \times p}$, $\mtx \vhbeta_{k}$ from Algorithm \ref{alg:lda-rk}, where the $i^{th}$ observation is selected at the $k^{th}$ step with probability $p_{i}=\frac{1}{n}$ satisfies
	\begin{eqnarray}\label{eqn:cor2}
		\Exp\left[\|\mtx\vhbeta_{k} - \mtx\vhbetals \|^{2}_{2} \given \mx\right]
		&\le& \left( 1 - 2c\Big(\frac{1}{\kappa(\mx)} - \frac{c}{n}\Big) \right)^{k} \Exp\|\mtx\|_{2}^{2} \, \|\vhbeta_{0} - \vhbetals\|^{2}_{2} \\
		&\;\;&+ \frac{\frac{c}{n}\kappa(\mx)}{1-\frac{c}{n}\kappa(\mx)} \, \kappa(\mx) \Exp\|\mtx\|_{2}^{2} \, r, \nonumber
	\end{eqnarray}
	where $r = \frac{\|\mx\vhbeta - \vy\|^{2}_{2}}{\|\mx\|^{2}_{F}}$.
	The conditional expectation is with respect to the sampling distribution of the iterates $\{i\}$ and the class-conditional distribution of $\mtx$.
\end{corollary}

Compared with \cite[Corollary 5.2]{needell2016stochastic}, which employs a diagonal scaling matrix $\md$ whose $i^{th}$ diagonal entry is $\|\vx_{i}\|_{2}$, Corollary \ref{cor:cor2} utilizes uniform sampling probabilities $p_{i} = \frac{1}{n}$.  This highlights the increased dependency on the conditioning of $\mx$: compared with Corollary \ref{cor:cor1}, Corollary \ref{cor:cor2} 
depends on $\kappa^{2}(\mx)$ rather than $\kappa(\mx)$.  If we employ a diagonal scaling matrix $\md$ as in \cite[Corollary 5.2]{needell2016stochastic}, we likewise observe exponential convergence to a radius of the weighted least squares solution scaled by $\Exp\|\mtx\|^{2}_{2}$.

Finally, since leverage score sampling is a common form of importance sampling in many sketching problems, Corollary \ref{cor:cor3} illustrates Theorem \ref{thm:expecteddiffXbeta} for \emph{leverage score sampling} with $p_{i} = \frac{\ell_{i}}{p}$.  While leverage score sampling is not recommended for massive unstructured matrices, our results hold for all data -- not only massive structured data.  Therefore, since leverage score sampling appears frequently in the matrix sketching literature, we include the following result to illustrate the flexibility of Theorem \ref{thm:expecteddiffXbeta}.

\begin{corollary}\label{cor:cor3}
	Given training data $\mx \in \mathbb{R}^{n \times (p+1)}$, where the first column of $\mx$ is $\vone_{n}$, and recoded class labels $\vy \in \mathbb{R}^{n}$ as in Algorithm \ref{alg:lda-rk}, let $\vhbeta$  be the minimizer of
	$
	\vhbeta = \arg \min_{\vbeta} \frac{1}{2} \| \mx \vbeta - \vy\|^{2}_{2},
	$
	and let $\vhbeta_{k}$ and $\vhbetals$ be the vectors formed from the last $p$ entries in $\vbeta_{k}$ from Algorithm \ref{alg:lda-rk} and from $\vhbeta$, respectively.  
	Let $c< \frac{p}{\kappa(\mx)}$ and let $\ell_{i}$ denote the $i^{th}$ row leverage score of the training data computed from the original $p$ variables only.  Then for any new unlabeled data $\mtx \in \mathbb{R}^{N \times p}$, $\mtx \vhbeta_{k}$ from Algorithm \ref{alg:lda-rk}, where the $i^{th}$ observation is selected at the $k^{th}$ step with probability $p_{i} = \frac{\ell_{i}}{p}$, satisfies
	\begin{eqnarray*}
		\Exp\left[\|\mtx\vhbeta_{k} - \mtx\vhbetals \|^{2}_{2} \given \mx\right]
		&\le& \left( 1 - 2c \Big(\frac{1}{\kappa(\mx)}-\frac{c}{p}\Big) \right)^{k}  \Exp\|\mtx\|_{2}^{2}\, \|\vhbeta_{0} - \vhbetals\|^{2}_{2} \\ &\;\;\;& +  \frac{\frac{c}{p} \, \kappa(\mx)}{1-\frac{c}{p}\,\kappa(\mx)}\kappa(\mx)\Exp\|\mtx\|_{2}^{2}\, r,
	\end{eqnarray*}
	where $r = \frac{\|\mx\vhbeta - \vy\|^{2}_{2}}{\|\mx\|^{2}_{F}}$. 
	The conditional expectation is with respect to the sampling distribution of the iterates $\{i\}$ and the class-conditional distribution of $\mtx$.
\end{corollary}

Here, we employ the fact that if $p_{i} = \frac{\ell_{i}}{p}$, then $\tilde{\alpha} = \frac{p}{\|(\mx\Tra\mx)\Dag\|_{2}}$ is a lower bound for the $\alpha_{i}$ to obtain the result.  Similar to Corollaries \ref{cor:cor1} and \ref{cor:cor2}, Corollary \ref{cor:cor3} does not depend on $n$.  However, the leverage score probabilities introduce a dependency on the number of variables $p$.  Finally, similar to Corollary \ref{cor:cor2}, Corollary \ref{cor:cor3} 
also depends on $\kappa^{2}(\mx)$ rather than $\kappa(\mx)$.

For notational simplicity, we do not distinguish whether the $p_{i}$ in Corollaries \ref{cor:cor1} and \ref{cor:cor3}  are computed with $\vone_{n}$ included in the training data $\mx$.  However, it appears reasonable to compute the $p_{i}$ with the original $p$ variables only.  Therefore, we adopt this approach in our numerical experiments in Section \ref{sec:experiments}.  Let $A = 2\frac{c }{\kappa(\mx)} \left(1-\frac{c}{\tilde{\alpha}} \|\mx\|^{2}_{F}\right)$ and $B = \frac{c}{\tilde{\alpha} (1-\frac{c}{\tilde{\alpha}})}$ from the convergence rate and horizon in Theorem \ref{thm:expecteddiffXbeta}, respectively.  Table \ref{table:results_with_sampling} summarizes differences in Corollaries \ref{cor:cor1} - \ref{cor:cor3}.

\begin{table}[h!]
	\centering
	\begin{tabular}{||c c c c||} 
		\hline
		Sampling approach & $p_{i}$ & A & B \\ [0.5ex] 
		\hline\hline 
		Row-weight sampling & $\|\vx_{i}\|^{2}_{2} / \|\mx\|^{2}_{F}$ & $2c\Big(\frac{1}{\kappa(\mx)}-\frac{c}{\kappa(\mx)}\Big)$ & $\frac{c}{1-c}$ \\ 
		Uniform sampling & $1 / n$ &  $2c\Big(\frac{1}{\kappa(\mx)} - \frac{c}{n}\Big)$ & $ \frac{\frac{c}{n}\kappa(\mx)}{1-\frac{c}{n}\kappa(\mx)} $ \\
		Leverage score sampling & $\ell_{i} / p$ & $2c \Big(\frac{1}{\kappa(\mx)}-\frac{c}{p}\Big)$ & $ \frac{\frac{c}{p} \, \kappa(\mx)}{1-\frac{c}{p}\,\kappa(\mx)}$ \\ [1ex] 
		\hline
	\end{tabular}
	\caption{\emph{Table summarizing differences in Corollaries \ref{cor:cor1} - \ref{cor:cor3}.}}
	\label{table:results_with_sampling}
\end{table}

\subsection{Change in optimal intercept due to randomized algorithm}
\label{sec:diffbetaopt}

We consider the change in $\betahopt$ from \eqref{eqn:optimalbetao} if one employs $\vhbetark$ in lieu of $\vhbetals$.  Denote this change by $\Delta \betahopt$ and observe that
\begin{eqnarray*}
	\Delta \betahopt &=& -\frac{1}{2}(\vhmu_{1} + \vhmu_{2})\Tra (\vhbetals-\vhbetark) \\ &&+ (\vhbetals - \vhbetark)\Tra \mhsig (\vhbetals-\vhbetark) \{(\vhmu_{2} - \vhmu_{1})\Tra (\vhbetals-\vhbetark) \}\Inv \log\left(\frac{n_{2}}{n_{1}}\right).
\end{eqnarray*}
Enveloping scalars inside Euclidean norms, applying the triangle and Cauchy-Schwarz inequalities, and the eigenvalue decomposition of $\mhsig$ produces
\begin{eqnarray*}
	\Delta \betahopt &\le& \frac{1}{2}\|\vhmu_{1} + \vhmu_{2}\| \|\vhbetals-\vhbetark\| + \frac{\hat{\sigma}_{1}^{2}(\mx) \|\vhbetals-\vhbetark\|^{2}} {\|(\vhmu_{2} - \vhmu_{1})\Tra (\vhbetals-\vhbetark) \|} \log\left(\frac{n_{2}}{n_{1}}\right),
\end{eqnarray*}
where $\hat{\sigma}_{1}(\mx)$ is the largest singular value of $\mx$. Applying the conditional expectation with respect to $\mx$ on both sides produces
\begin{eqnarray}\label{eqn:changeinbetaopt}
	\Exp [\Delta \betahopt \given \mx] &\le& \frac{1}{2}\|\vhmu_{1} + \vhmu_{2}\| \Exp[\|\vhbetals-\vhbetark\|\given \mx]  \\
	&&+ \hat{\sigma}_{1}^{2}(\mx)\log\left(\frac{n_{2}}{n_{1}}\right)\Exp\left[\frac{ \|\vhbetals-\vhbetark\|^{2}} {\|(\vhmu_{2} - \vhmu_{1})\Tra (\vhbetals-\vhbetark) \|}\Big| \, \mx \right] .\nonumber
\end{eqnarray}
Theorem 2.1 from \cite{needell2016stochastic} shows that the first term on the right in \eqref{eqn:changeinbetaopt} can be made small with sufficiently many $k$ for any given tolerance $\epsilon>0$ and suitably chosen $p_{i}$.  For this reason, we likewise expect the numerator in the expectation of the second term to decrease faster than the denominator.  Of course, this requires that the class means are sufficiently separated so that the data are suitable for LDA and that $(\vhmu_{2} - \vhmu_{1})$ and $(\vhbetals-\vhbetark)$ are not orthogonal.  Therefore, under some mild assumptions, it appears that the expected change in the optimal intercept from \eqref{eqn:optimalbetao} conditioned on $\mx$ can likewise be made suitably small.  

\section{Proofs of Main Results}
\label{sec:proofs}

In this section, we present proofs of the main results.  We begin by restating \cite[Lemma 8.1]{needell2016stochastic}.
\begin{lemma}[Definition 2.1 from \cite{zhu1996co}, Lemma 8.1 from \cite{needell2016stochastic}(Co-coercivity)]\label{lemma:cocoercivity}
	For a smooth function $f$ whose gradient has Lipschitz constant L, 
	\begin{eqnarray*}
		\| \nabla f(\vx) - \nabla f(\vy) \|^{2}_{2} \le L \langle \vx - \vy, \nabla f(\vx) - \nabla f(\vy) \rangle.
	\end{eqnarray*}
\end{lemma}

\subsection{Proof of Theorem \ref{thm:expecteddiffXbeta}}

We first present a general result in the SGD framework.  We then tie this result to the RK by updating the SGD components via the reweighted SGD framework and the corresponding sampling weights as in \cite{needell2016stochastic}.
Following the notation in \cite[Section 2]{needell2016stochastic}, we seek to minimize a strongly convex function $F$, where $F(\vbeta) = \Exp_{i\sim \mathcal{D}} f_{i}(\vbeta)$ and each $f_{i}: \mathbb{R}^{d} \rightarrow \mathbb{R}$ is convex with Lipschitz constant $\inf L \le L_{i} \le \sup L$ almost surely.  Here, the expectation is with respect to the sampling distribution of the SGD iterates $\{i\}$ according to some source distribution $\mathcal{D}$, where the $i^{th}$ row of $\mx$ is selected at the $k^{th}$ iterate.  Let $\bar{L}$ denote the average $L_{i}$.  Let $\sigma^{2} = \Exp \|\nabla f_{i}(\vbetastar) \given \mx, \mtx \|^{2}_{2}$ denote the expected norm of the residual gradient at the minimizer $\vbetastar = \arg \min_{\vbeta} F(\vbeta)$ conditioned on the data $\mx$ and $\mtx$, and let $\gamma < \frac{1}{\sup L}$ be a fixed step-size.  

For any new unlabeled data matrix $\mtx \in \mathbb{R}^{N \times (p+1)}$ and the $k+1^{th}$ iterate, we employ the definition of the operator norm and the SGD iterate update rule
$
\vbeta_{k+1} \leftarrow \vbeta_{k} - \gamma \nabla f_{i}(\vbeta_{k}) 
$
\, to obtain
\begin{eqnarray}\label{eqn:sgd-iterates-1}
	\|\mtx\vbeta_{k+1} - \mtx\vbetastar \|^{2}_{2} &\le& \| \mtx \|_{2}^{2} \, \| \vbeta_{k} - \vbetastar - \gamma \nabla f_{i}(\vbeta_{k}) \|^{2}_{2} \nonumber \\
	&=& \| \mtx\|_{2}^{2} \Big(\| \vbeta_{k} - \vbetastar \|^{2}_{2} - 2\gamma \langle \vbeta_{k} - \vbetastar, \nabla f_{i}(\vbeta_{k}) \rangle + \gamma^{2} \| \nabla f_{i}(\vbeta_{k}) \|^{2}_{2}\Big).
\end{eqnarray}
Focusing on the last term in \eqref{eqn:sgd-iterates-1}, we add and subtract $\nabla f_{i}(\vbetastar)$ inside the norm, scale terms inside the norm by $2 \cdot \frac{1}{2}$, and simplify to obtain
\begin{eqnarray}\label{eqn:sgd-iterates-2}
	\gamma^{2}\| \nabla f_{i}(\vbeta_{k}) \|^{2}_{2} &=&
	4\gamma^{2}\left\| \frac{1}{2}\Big(\nabla f_{i}(\vbeta_{k}) - \nabla f_{i}(\vbetastar)\Big) + \frac{1}{2} \nabla f_{i}(\vbetastar) \right\|^{2}_{2} \nonumber \\
	&\le& 2\gamma^{2}\| \nabla f_{i}(\vbeta_{k}) - \nabla f_{i}(\vbetastar)\|^{2}_{2} + 2 \gamma^{2} \| \nabla f_{i}(\vbetastar) \|^{2}_{2},
\end{eqnarray}
where we apply Jensen's inequality to obtain the inequality in the second line.
Employing \eqref{eqn:sgd-iterates-2} in \eqref{eqn:sgd-iterates-1} then produces
\begin{eqnarray*}
	\|\mtx\vbeta_{k+1} - \mtx\vbetastar \|^{2}_{2} 
	&\le& \| \mtx \|_{2}^{2} \Big( \| \vbeta_{k} - \vbetastar \|^{2}_{2} - 2\gamma \langle \vbeta_{k} - \vbetastar, \nabla f_{i}(\vbeta_{k}) \rangle \\
	&&+ 2\gamma^{2}\| \nabla f_{i}(\vbeta_{k}) - \nabla f_{i}(\vbetastar)\|^{2}_{2} + 2 \gamma^{2} \| \nabla f_{i}(\vbetastar) \|^{2}_{2} \Big).
\end{eqnarray*}
Applying Lemma \ref{lemma:cocoercivity} to $\| \nabla f_{i}(\vbeta_{k}) - \nabla f_{i}(\vbetastar)\|^{2}_{2}$ then gives us
\begin{eqnarray*}
	\|\mtx\vbeta_{k+1} - \mtx\vbetastar \|^{2}_{2} 
	&\le& \|\mtx\|_{2}^{2} \| \vbeta_{k} - \vbetastar \|^{2}_{2} - 2\gamma \|\mtx\|_{2}^{2}\langle \vbeta_{k} - \vbetastar, \nabla f_{i}(\vbeta_{k}) \rangle \\
	&& + 2\gamma^{2} \|\mtx\|_{2}^{2} \,L_{i}\, \langle \vbeta_{k} - \vbetastar, \nabla f_{i}(\vbeta_{k}) - \nabla f_{i}(\vbetastar) \rangle + 2 \gamma^{2} \|\mtx\|_{2}^{2}\|\nabla f_{i}(\vbetastar) \|^{2}_{2}.
\end{eqnarray*}

Employing the law of iterated expectations, we first take the expectation on both sides with respect to the sampling distribution of $\{ i\}$ while conditioning on both the training data $\mx$ and the new data $\mtx$.  We employ the facts that ${F(\vbeta) = \Exp_{i\sim\mathcal{D}} [f_{i}(\vbeta)\given \mx, \mtx ]}$, 
$\Exp_{i\sim\mathcal{D}} [\nabla f_{i}(\vbeta)\given \mx, \mtx ] = \nabla F(\vbeta)$, ${\sigma^{2} = \Exp_{i\sim\mathcal{D}} \|\nabla f_{i}(\vbetastar)\given \mx, \mtx \|^{2}}$, and $L_{i} \le \sup L$ almost surely to obtain
\begin{eqnarray*}
	\Exp_{i\sim\mathcal{D}}\left[\|\mtx\vbeta_{k+1} - \mtx\vbetastar \|^{2}_{2} \given \mx, \mtx\right] 
	&\le& \|\mtx\|_{2}^{2}\| \vbeta_{k} - \vbetastar \|^{2}_{2} - 2\gamma \|\mtx\|_{2}^{2} \langle \vbeta_{k} - \vbetastar, \nabla F(\vbeta_{k}) \rangle \\
	&& + 2\gamma^{2}\|\mtx\|^{2}_{2} \,\sup L \, \langle \vbeta_{k} - \vbetastar, \nabla F(\vbeta_{k}) - \nabla F(\vbetastar) \rangle + 2 \gamma^{2} \|\mtx\|^{2}_{2}\sigma^{2}.
\end{eqnarray*}
Since $\vbetastar$ minimizes $F$, we have $\nabla F(\vbetastar) = 0$ so that
\begin{eqnarray}\label{eqn:exp-sgd-iterates}
	\Exp_{i\sim\mathcal{D}}\left[\|\mtx\vbeta_{k+1} - \mtx\vbetastar \|^{2}_{2} \given \mx, \mtx\right]
	&\le& 
	- 2\gamma \|\mtx\|_{2}^{2}(1-\gamma \sup L) \langle \vbeta_{k} - \vbetastar, \nabla F(\vbeta_{k}) \nonumber  \\ &&- \nabla F(\vbetastar)\rangle 
	+\|\mtx\|_{2}^{2}\| \vbeta_{k} - \vbetastar \|^{2}_{2} + 2 \gamma^{2} \|\mtx\|^{2}_{2}\sigma^{2}. 
\end{eqnarray}
We employ the fact that $F$ is $\mu$-strongly convex so that for $\mu > 0$ we have
\begin{eqnarray}\label{eqn:stronglyconvex}
	-\langle \vbeta_{k} - \vbetastar, \nabla F(\vbeta_{k}) - \nabla F(\vbetastar)\rangle \le -\mu \|\vbeta_{k} - \vbetastar\|^{2}_{2}.
\end{eqnarray}
Therefore, utilizing \eqref{eqn:stronglyconvex} in \eqref{eqn:exp-sgd-iterates} gives us
\begin{eqnarray*}
	\Exp_{i\sim\mathcal{D}}\left[\|\mtx\vbeta_{k+1} - \mtx\vbetastar \|^{2}_{2} \given \mx, \mtx\right]
	&\le& \|\mtx\|_{2}^{2}\left[1 - 2\gamma \mu (1-\gamma \sup L)\right] \|\vbeta_{k} - \vbetastar\|^{2}_{2} + 2 \gamma^{2} \|\mtx\|^{2}_{2}\sigma^{2}
\end{eqnarray*}
since $\gamma \le \frac{1}{\sup L}$.  Now taking the expectation on both sides with respect to the class-conditional distribution of the new data $\mtx$ gives us
\begin{eqnarray*}
	\Exp_{\mtx}\left[\|\mtx\vbeta_{k+1} - \mtx\vbetastar \|^{2}_{2} \given \mx\right]
	&\le& \left[1 - 2\gamma \mu (1-\gamma \sup L)\right] \|\vbeta_{k} - \vbetastar\|^{2}_{2}\Exp\|\mtx\|_{2}^{2} + 2 \gamma^{2}\sigma^{2} \Exp\|\mtx\|^{2}_{2}.
\end{eqnarray*}		
Since we employed $
\|\mtx\vbeta_{k+1} - \mtx\vbetastar \|^{2}_{2} 
\le \|\mtx\|^{2}_{2} \, \|\vbeta_{k} - \vbetastar \|^{2}_{2}
$
in \eqref{eqn:sgd-iterates-1}, it follows that
\begin{eqnarray*}
	\Exp_{\mtx}\left[\|\mtx\vbeta_{k+1} - \mtx\vbetastar \|^{2}_{2} \given \mx\right] 
	&\le& \Exp_{\mtx}\left[\|\mtx\|^{2}_{2} \, \|\vbeta_{k} - \vbetastar \|^{2}_{2} \given \mx\right] \\ 
	&\le& \left[1 - 2\gamma \mu (1-\gamma \sup L)\right] \|\vbeta_{k} - \vbetastar\|^{2}_{2}\Exp\|\mtx\|_{2}^{2} + 2 \gamma^{2}\sigma^{2} \Exp\|\mtx\|^{2}_{2}
\end{eqnarray*} above.  After dividing both sides by $\Exp\|\mtx\|^{2}_{2}$, we recursively apply this bound over the previous $k$ iterations to obtain 
\begin{eqnarray}
	\frac{\Exp\left[\|\mtx\vbeta_{k} - \mtx\vbetastar \|^{2}_{2} \given \mx\right]}{\Exp\|\mtx\|^{2}_{2}}
	&\le& \Big( 1 - 2\gamma \mu (1-\gamma \sup L) \Big)^{k} \|\vbeta_{0} - \vbetastar\|^{2}_{2} \nonumber \\
	&&+ 2 \sum_{j=0}^{k-1} \Big( 1 - 2\gamma \mu (1-\gamma \sup L) \Big)^{j} \gamma^{2}\sigma^{2}  \nonumber \\
	&\le& \Big( 1 - 2\gamma \mu (1-\gamma \sup L) \Big)^{k} \|\vbeta_{0} - \vbetastar\|^{2}_{2} + \frac{\gamma \sigma^{2}}{\mu(1-\gamma \sup L)}.
\end{eqnarray}
This gives us the following general SGD result
\begin{eqnarray}\label{eqn:genresult}
	\Exp\left[\|\mtx\vbeta_{k} - \mtx\vbetastar \|^{2}_{2} \given \mx\right]
	&\le& \Big( 1 - 2\gamma \mu (1-\gamma \sup L) \Big)^{k} \Exp\|\mtx\|^{2}_{2} 
	\; \|\vbeta_{0} - \vbetastar\|^{2}_{2} + \nonumber \\
	&\;\;& \frac{\gamma \sigma^{2}}{\mu(1-\gamma \sup L)}\Exp\|\mtx\|^{2}_{2}.
\end{eqnarray}

We update this general result for the least squares problem via randomized Kaczmarz with importance sampling \cite[Section 5]{needell2016stochastic}.  
In particular, we first begin with the unweighted SGD problem
$
\arg \min_{\vbeta} F(\vbeta) = \arg \min_{\vbeta} \frac{1}{2} \| \mx \vbeta - \vy\|^{2}_{2},
$ 
where $F$ is $\mu$-strongly convex with $\mu = \frac{1}{\|(\mx\Tra\mx)\Inv\|_{2}}$ and the solution is $\vhbeta = \arg \min_{\vbeta} F(\vbeta)$. 
Therefore, $f_{i} = \frac{n}{2}(\vx_{i}\Tra\vbeta - y_{i})^{2}$ is convex with Lipschitz constant $L_{i} = n\|\vx_{i}\|^{2}_{2}$, and $\nabla f_{i} = n(\vx_{i}\Tra\vbeta - y_{i})\vx_{i}$.
Let $\alpha_{i} = \frac{\|\vx_{i}\|^{2}_{2}}{p_{i}}$ and let $\tilde{\alpha}$ be a lower bound for $\alpha_{i}$ so that $\tilde{\alpha} \le \alpha_{i} \le \sup_{i} \alpha_{i}$ almost surely.  Let $\bar{L} = \Exp_{i}[L_{i}]$.  Then for any normalized sampling weight $w_{i} = n \cdot p_{i}$, we have reweighted SGD components \cite[Sections 3.2 and 3.3]{needell2016stochastic}
\begin{eqnarray*} 
	&&\fiw(\vbeta) = \frac{1}{w_{i}}f_{i}(\vbeta), \quad
	F^{(w)}(\vbeta) = \Exp^{(w)}[f_{i}^{(w)}(\vbeta)], \quad \\
	&&L_{i}^{(w)} = \frac{1}{w_{i}}L_{i} = \frac{\|\vx_{i}\|^{2}_{2}}{p_{i}} = \alpha_{i}, \quad
	\sup_{i} L^{(w)}_{i} = \sup_{i} \frac{1}{w_{i}}L_{i} = \bar{L} = \|\mx\|^{2}_{F} = \sup_{i} \alpha_{i}, \text{ and} \\ 
	&&\sigma^{2}_{(w)} = \Exp\Big[\frac{1}{w_{i}}\| \nabla f_{i}(\vhbetals) \|^{2}_{2}\Big] = \sum_{i=1}^{n} \frac{\|\vx_{i}\|^{2}_{2}}{p_{i}} \,| \vx_{i}\Tra\vhbetals - y_{i}|^{2} \le \sup_{i} \alpha_{i} \, \|\mx\vhbetals - \vy\|^{2}_{2} ,
\end{eqnarray*}
and reweighted SGD iterates 
$
\vbeta_{k+1} \leftarrow \vbeta_{k} -\frac{\gamma}{w_{i}}\nabla f_{i}(\vbeta_{k}).
$
We rewrite the SGD step-size parameter $\gamma$ in terms of the step-size parameter $c$ for the randomized Kaczmarz method in \eqref{eqn:rk} with $\gamma = \frac{c}{\alpha_{i}}$.  We then update \eqref{eqn:genresult} with the reweighted SGD components $L_{i}^{(w)}, \sup_{i} L^{(w)}_{i},$ and $\sigma^{2}_{(w)}$ with fixed step-size $c < \frac{\alpha_{i}}{\sup_{i} \alpha_{i}} \le 1$ to obtain 
\begin{eqnarray*}
	\Exp\left[\|\mtx\vbeta_{k} - \mtx\vhbeta \|^{2}_{2} \given \mx\right]
	&\le& \left( 1 - 2\frac{c \mu}{\alpha_{i}} \Big(1-\frac{c}{\alpha_{i}} \|\mx\|^{2}_{F}\Big) \right)^{k} \Exp\|\mtx\|_{2}^{2} \; \|\vbeta_{0} - \vhbeta\|^{2}_{2} + \frac{c}{\alpha_{i}}\, \frac{\|\mx\|^{2}_{F} \Exp\|\mtx\|_{2}^{2} \,r^{\star}}{\mu(1-\frac{c}{\alpha_{i}} \|\mx\|^{2}_{F})}, \\
	&\le& \left( 1 - 2\frac{c }{\kappa(\mx)} \Big(1-\frac{c}{\tilde{\alpha}} \|\mx\|^{2}_{F}\Big) \right)^{k} \Exp\|\mtx\|_{2}^{2} \; \|\vbeta_{0} - \vhbeta\|^{2}_{2} + \frac{c}{\tilde{\alpha}}\, \frac{\kappa(\mx) \Exp\|\mtx\|_{2}^{2}\,r^{\star}}{(1-\frac{c}{\tilde{\alpha}} \|\mx\|^{2}_{F})}, \nonumber
\end{eqnarray*}
where $r^{\star} = \|\mx\vhbeta - \vy\|^{2}_{2}$.
Finally, let $\vhbeta_{k}$ and $\vhbetals$ be the vectors formed from the last $p$ entries in $\vbeta_{k}$ from Algorithm \ref{alg:lda-rk} and $\vhbeta$, respectively.  Since we have an upper bound on the squared Euclidean norm of the $p+1$ entries in $\Exp\left[\|\mtx\vbeta_{k} - \mtx\vhbeta \|^{2}_{2} \given \mx\right]$, the same right hand side is also an upper bound on the $p$ entries in $\Exp\left[\|\mtx\vhbeta_{k} - \mtx\vhbetals \|^{2}_{2} \given \mx\right]$.$\Box$

\subsection{Proof of Corollary \ref{cor:cor-iterationsk}}

We set the total error to $\epsilon \cdot (1+r^{\star})$ and employ $c = \frac{\epsilon \, \tilde{\alpha}}{2 (\kappa(\mx) \Exp\|\mtx\|_{2}^{2}  + \epsilon \|\mx\|^{2}_{F})}$ in the second term on the right of \eqref{eqn:mainresult} to obtain
\begin{eqnarray*}
	\frac{c}{\tilde{\alpha}}\, \frac{\kappa(\mx) \Exp\|\mtx\|_{2}^{2}}{(1-\frac{c}{\tilde{\alpha}} \|\mx\|^{2}_{F})} \le \frac{\epsilon}{2}.
\end{eqnarray*}
Therefore, inserting this $c$ in the first term on the right of \eqref{eqn:mainresult}, requiring that 
\begin{eqnarray*}
	\left( 1 - 2\frac{c }{\kappa(\mx)} \Big(1-\frac{c}{\tilde{\alpha}} \|\mx\|^{2}_{F}\Big) \right)^{k} \Exp\|\mtx\|_{2}^{2} \; \epsilon_{0} \le \frac{\epsilon}{2},
\end{eqnarray*}
and rearranging shows us that we need a $k$ that satisfies
\begin{eqnarray*}
	k \, \log ( 1 - A ) \le - \log\left( \frac{2 \epsilon_{0} \Exp\|\mtx\|^{2}_{2} }{\epsilon} \right),
\end{eqnarray*}
where $A = \frac{4\epsilon \Exp\|\mtx\|^{2}_{2}}{\epsilon \|\mx\|^{2}_{F} + 2 \kappa(\mx)\Exp\|\mtx\|^{2}_{2}}$.  Here, we use the fact that  $\inf_{i} \alpha_{i}\le \sup_{i} \alpha_{i} = \|\mx\|^{2}_{F}$ almost surely.  Supposing that $\mx$ and $\mtx$ are not all zeros matrices, $A > 0$.  Also, $A \le 1$ for any $\epsilon < \frac{1}{2}$.  Therefore, we can always choose an $\epsilon$ small enough so that $0 < A \le 1$.  Using the fact that for any $0 < A \le 1$, $-\frac{1}{\log(1-A)} \le \frac{1}{A}$ and rearranging gives us
\begin{eqnarray*}
	k \ge \log\Big( \frac{2 \epsilon_{0} \Exp\|\mtx\|^{2}_{2} }{\epsilon} \Big) 
	\left( \frac{\|\mx\|^{2}_{F}}{4\Exp\|\mtx\|^{2}_{2}} + \frac{\kappa(\mx)}{2\epsilon} \right).
\end{eqnarray*}

\section{Numerical Experiments}
\label{sec:experiments}

We perform numerical experiments with rkLDA from Algorithm \ref{alg:lda-rk} on real datasets with known labels to investigate several questions. (1) How does rkLDA compare with full-data LDA from \eqref{eqn:discriminantfuncs-v3} (gmLDA) and full data least squares LDA from Algorithm \ref{alg:lda-ls} (lsLDA)?  (2) How does performance vary with number of iterations and step-sizes?  (3) How does performance vary with smaller and larger values of $p$?  (4) How does rkLDA compare with randomized matrix sketching (see Section \ref{sec:relatedworks}) and least squares solvers such as CG applied to the normal equations?

To investigate question (3), we perform two sets of experiments on datasets with varying number of features $p$.  (i)The first employs two datasets with a small number of features $p$ and investigates questions (1) and (2).  In particular, we explore how the gmLDA and lsLDA classifiers are aligned, and whether or not the rkLDA iterates can obtain a solution that is close to the gmLDA classifier.  (ii) The second set employs two datasets with much larger $p$ and investigates questions (2) and (4) on larger data, and with comparisons to randomized matrix sketching and CG.  Since the solution vectors from larger $p$ datasets are high-dimensional we do not examine angles in those experiments.  Instead, we investigate the change in the LDA discriminant function when employing the RK solution at the $k^{th}$ iterate in lieu of the full data LS solution $\| \mtx\vhbeta_{k} - \mtx\vhbetals \|$.  Notably, our results demonstrate that rkLDA can achieve comparable classification accuracy to gmLDA even before the solution vectors are closely aligned.

\subsection{Experimental Setup}

In the first set of experiments, our comparison metrics include the angle (in degrees) from gmLDA and classification accuracy.  We compute the angle without the intercept since gmLDA does not employ one.  Whenever possible, we obtain the classifier for gmLDA with the \verb|lda| function in the \verb|MASS| package for R since it is the defacto method for performing LDA in R.  In cases where the LDA classifier from \verb|MASS| does not coincide with the gmLDA classifier from \eqref{eqn:discriminantfuncs-v3}, we instead employ \eqref{eqn:discriminantfuncs-v3}.  For each scenario, we present the mean of the comparison metric over 20 replicates for each combination of step-size $c$ and number of iterations $k$.  Values for $c$ are $0.1, 0.3, 0.5, 0.7$, and $0.9$.  Values for $k$ vary by dataset.  

In the second set of experiments, our comparison metrics include classification accuracy and time (in seconds).  We compute the gmLDA classifier using the \verb|LinearDiscriminantAnalysis| module from the Scikit-learn package for Python \cite{scikit-learn}.  For each scenario, we present boxplots of the comparison metric over $100$ replicates for each combination of step-size $c$ and number of iterations $k$.  Values for $c$ are $0.1, 0.3, 0.5, 0.7$, and $0.9$.  Values $k$ are $500, 1{,}000, 1{,}500, 2{,}000$, and $2{,}500$.

The second experiments include comparisons to randomized matrix sketching (MSLDA) and CG.  For sketching experiments, we sample the rows of both the design and response for comparability to rkLDA and sample $k$ rows with replacement in each replicate.  We do not sample the columns since rkLDA does not perform dimension reduction on the feature space.  Additionally, we do not compare other forms of sketching matrices (e.g. Gaussian or FJLT sketching) in order to preserve the values in the response for classification.  To obtain the best timing results for sketching, we do not form the sampling matrices and merely subset the selected observations in each replicate.  We employ the \verb|scipy.sparse.linalg.cg| module from the Scipy package for Python \cite{2020SciPy-NMeth} to perform CG on the normal equations.  We additionally examine the change in the LDA discriminant function when employing the RK solution at the $k^{th}$ iterate in lieu of the full data LS solution $\| \mtx\vhbeta_{k} - \mtx\vhbetals \|$ for varying values of $k$ and step-sizes $c$ over $50$ replicates.  All experiments were run on an Intel Core i7 3.00 GHz machine with 32 GB of RAM.

\subsection{Experiments illustrating angle and accuracy}

We present data and results from experiments illustrating angle and accuracy.

\subsubsection{Mammographic Mass Data}

We employ the mammographic mass dataset \cite{elter2007prediction} containing $830$ observations with complete measurements on four features: patient age, mass shape, mass margin, and mass density.  The goal is to predict whether a mass is benign or malignant given the observed features.  We randomly split the data into training (80 percent) and testing (20 percent) sets and perform 20 replicates of rkLDA on this split so that differences among the replicates are due solely to algorithmic randomness in Algorithm \ref{alg:lda-rk}.  The training and testing datasets have rank $4$ and scaled condition numbers equal to $7{,}505$ and $7{,}312$, respectively.

Table \ref{tab:mammographic} depicts comparisons between gmLDA, lsLDA, and rkLDA.  The top portion reports the coefficients of classifiers from gmLDA, lsLDA, and a single replicate of rkLDA with $c=0.9$ and $k=1{,}000{,}000$ iterations with the least squares (LS) and optimal intercepts (opt) indicated in subscript.  The middle reports the angles of the lsLDA and rkLDA solutions from the gmLDA classifier.  
The bottom reports overall and class-specific classification accuracy (with classes denoted in subscript) including the intercept whenever possible.  Values for the number of iterations $k$ come from an equally spaced sequence of ten numbers from $10^{3}$ to $10^{6.5}$.  Therefore, $k$ ranges from $1000$ to approximately $3{,}000{,}000$.

\begin{table}[ht]
	\centering
	\resizebox{\columnwidth}{!}{%
		\begin{tabular}{lrrrrrrr}
			\hline
			& gmLDA & lsLDA$_{\text{LS}}$ & lsLDA$_{\text{opt}}$ & Scaling & rkLDA$_{\text{LS}}$ & rkLDA$_{\text{opt}}$ & Scaling \\ 
			\hline
			Intercept &  & -3.17 & -3.17 &  & -3.05 &-3.49 &  \\ 
			Age & 0.02 & 0.02 & 0.02 & 0.99 & 0.03 & 0.03 & 0.93 \\ 
			Shape & 0.50 & 0.51 & 0.51 & 0.99 & 0.54 & 0.54 & 0.94 \\ 
			Margin & 0.40 & 0.40 & 0.40 & 0.99 & 0.40 & 0.40 & 1.00 \\ 
			Density & -0.20 & -0.20 & -0.20 & 0.99 & -0.25 & -0.25 & 0.82 \\ 
			\hline
			Angle &  & 0.00 &  &  & 3.35 &  &  \\ 
			\hline
			Accuracy & {\bf 0.80} & 0.79 & 0.79 &  & 0.79 & {\bf 0.80} &  \\ 
			Accuracy$_{1}$ & 0.74 & 0.71 & 0.75 &  & 0.69 & {\bf 0.76} &  \\ 
			Accuracy$_{2}$ & 0.85 & {\bf 0.88} & 0.84 &  & {\bf 0.88} & 0.84 &  \\ 
			\hline 
		\end{tabular}
	}
	\caption{\begin{footnotesize}\label{tab:mammographic}\emph{Top: Coefficients from gmLDA, lsLDA, and rkLDA classifiers with least squares (LS) and optimal (opt) intercepts denoted in subscript.  Middle: Angle (degrees) from gmLDA classifier.  Bottom: Overall classification accuracy and class-specific accuracy.}\end{footnotesize} }
\end{table}

The lsLDA and gmLDA coefficients are the same up to a scaling factor of $0.99$.  Therefore, gmLDA and lsLDA produce the same classifiers and the angle between them is $0$.  Meanwhile, the rkLDA solution after $k=1{,}000{,}000$ iterations is $3.35$ degrees from the gmLDA classifier.  The lsLDA and rkLDA perform similarly to gmLDA on overall accuracy with some variation in the class-specific accuracy (Accuracy$_{1}$ and Accuracy$_{2}$ in Table \ref{tab:mammographic}).

\begin{figure}[ht]
	\centering
	\begin{subfigure}{.47\textwidth}
		\centering
		\includegraphics[width=\linewidth]{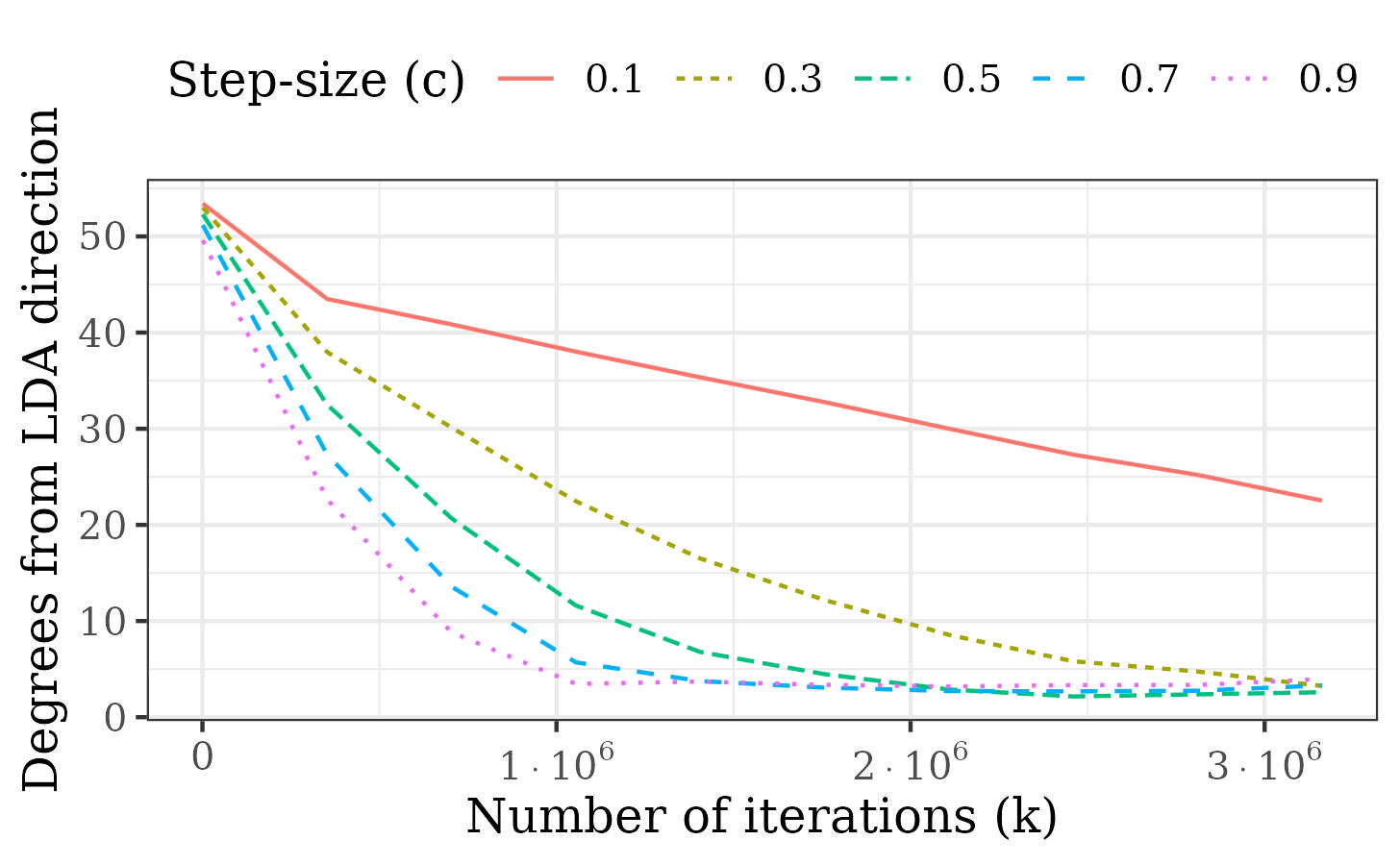}
		\caption{\begin{footnotesize}\label{fig:mammographic_angle}\emph{Mean angle (degrees) between LDA and rkLDA classifiers.}\end{footnotesize} }
	\end{subfigure}%
	\hspace{0.01\linewidth}
	\begin{subfigure}{.47\textwidth}
		\centering
		\includegraphics[width=\linewidth]{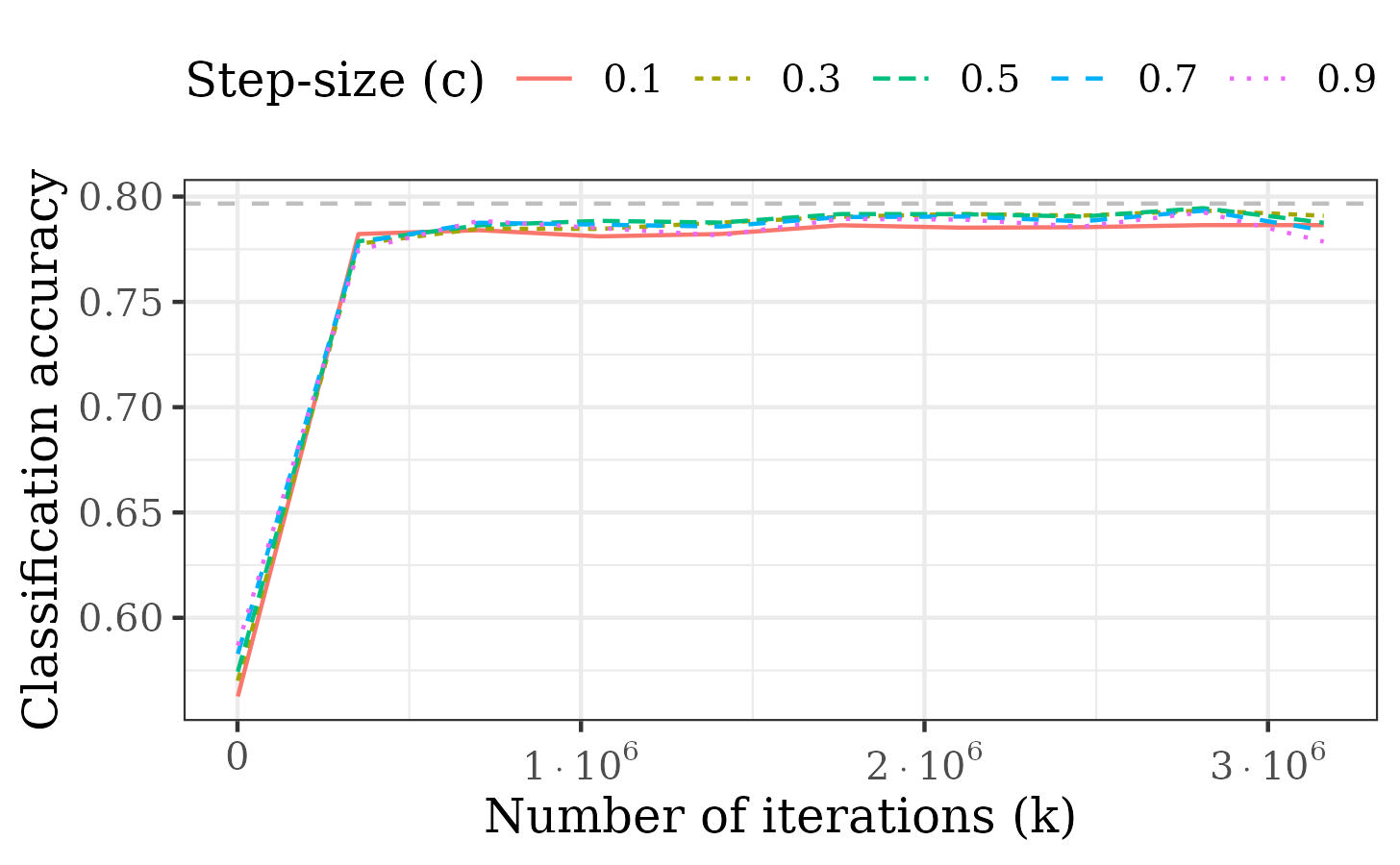}
		\caption{\begin{footnotesize}\label{fig:mammographic_accuracy}\emph{Mean rkLDA accuracy.  Gray line depicts gmLDA accuracy ($0.797$).}\end{footnotesize} }
	\end{subfigure}
	\vspace{-.6\baselineskip}
	\caption{\begin{footnotesize}\emph{Mammographic mass results over 20 replicates as a function of iterations.}\end{footnotesize}}
	\label{fig:mammographicmass}
\end{figure}

Figure \ref{fig:mammographic_angle} depicts the mean angle between the gmLDA and rkLDA classifiers obtained over 20 replicates as a function of the number of iterations $k$.  Colored lines depict results from different step-size values $c$.  While results with all step-sizes appear to approach the gmLDA classifier, larger step-sizes converge more quickly to a solution that is very close to the gmLDA classifier.  
Figure \ref{fig:mammographic_accuracy} depicts the mean overall classification accuracy of the rkLDA classifier obtained over 20 replicates with the optimal intercept as a function of the number of iterations $k$.  The dashed gray line depicts the gmLDA overall classification accuracy.  Notably, accuracy from all choices of step-size $c$ approach the gmLDA accuracy even before the angle from the gmLDA classifier becomes very small.  

\subsubsection{Occupancy Detection Data}

We employ the occupancy detection dataset from \cite{candanedo2016accurate} containing $8{,}143$ training and $9{,}752$ testing observations on four features: temperature (in Celsius), relative humidity (percent), light (in lux), and CO$_{2}$ (in ppm).  The goal is to predict whether or not a room is occupied given the observed features.  We perform 20 replicates of rkLDA on the training data so that differences among the replicates are due solely to algorithmic randomness from the sampling procedure in Algorithm \ref{alg:lda-rk}.  The training and testing datasets have rank $4$ and scaled condition numbers equal to $43{,}979$ and $120{,}589$, respectively.

Table \ref{tab:occupancy} depicts comparisons between gmLDA, lsLDA, and rkLDA.  The top portion reports the coefficients of the classifiers from gmLDA, lsLDA, and a single iterate of rkLDA with $c=0.9$ and $k=100{,}000$ iterations with the least squares (LS) and optimal intercepts (opt) indicated in subscript.  The middle reports the angles of the lsLDA and rkLDS solutions from the gmLDA classifier.  
The bottom reports the overall and class-specific classification accuracy (with classes denoted in subscript) including the intercept whenever possible.  Values for the number of iterations $k$ come from an equally spaced sequence of ten numbers from $10^{3}$ to $10^{6}$.  Therefore, $k$ ranges from $1000$ to approximately $1{,}000{,}000$.

\begin{table}[! h]
	\centering
	\resizebox{\columnwidth}{!}{%
		\begin{tabular}{lrrrrrrr}
			\hline
			& gmLDA & lsLDA$_{\text{LS}}$ & lsLDA$_{\text{opt}}$ & Scaling & rkLDA$_{\text{LS}}$ & rkLDA$_{\text{opt}}$ & Scaling \\ 
			\hline
			Intercept &  & 5.65 & 2.86 &  & 0.94 & -2.28 &  \\ 
			Temperature & -0.44 & -0.38 & -0.38 & 1.17 & -0.14 & -0.14 & 3.17 \\ 
			Humidity & -0.02 & -0.01 & -0.01 & 1.17 & -0.01 & -0.01 & 1.35 \\ 
			Light & 0.01 & 0.01 & 0.01 & 1.17 & 0.01 & 0.01 & 0.99 \\ 
			CO$_{2}$ & 0.00 & 0.00 & 0.00 & 1.17 & 0.00 & 0.00 & 1.78 \\ 
			\hline
			Angle &  & 0.00 &  &  & 4.63 &  &  \\ 
			\hline
			Accuracy & {\bf 0.99} & 0.88 & 0.98 &  & 0.94 & {\bf 0.99} &  \\ 
			Accuracy$_{1}$ & {\bf 0.99} & 0.85 & {\bf 0.99} &  & 0.93 & {\bf 0.99} &  \\ 
			Accuracy$_{2}$ & {\bf 1.00} & {\bf 1.00} & 0.93 &  & {\bf 1.00} & 0.99 &  \\ 
			\hline 
		\end{tabular}
	}
	\caption{\begin{footnotesize}\label{tab:occupancy}\emph{Top: Coefficients from gmLDA, lsLDA, and rkLDA classifiers with least squares (LS) and optimal (opt) intercepts denoted in subscript.  Middle: Angle (degrees) from gmLDA classifier.  Bottom: Overall classification accuracy and class-specific accuracy.}\end{footnotesize} }
\end{table}

The lsLDA and gmLDA coefficients are the same up to a scaling factor of $1.17$.  Therefore, gmLDA and lsLDA produce the same classifiers and the angle between them is $0$.  Meanwhile, the rkLDA solution after $k=100{,}000$ iterations is $4.63$ degrees from the gmLDA classifier.  Both lsLDA and rkLDA produce overall accuracy that is very comparable to that of gmLDA.

\begin{figure}[ht]
	\centering
	\begin{subfigure}{.47\textwidth}
		\centering
		\includegraphics[width=\linewidth]{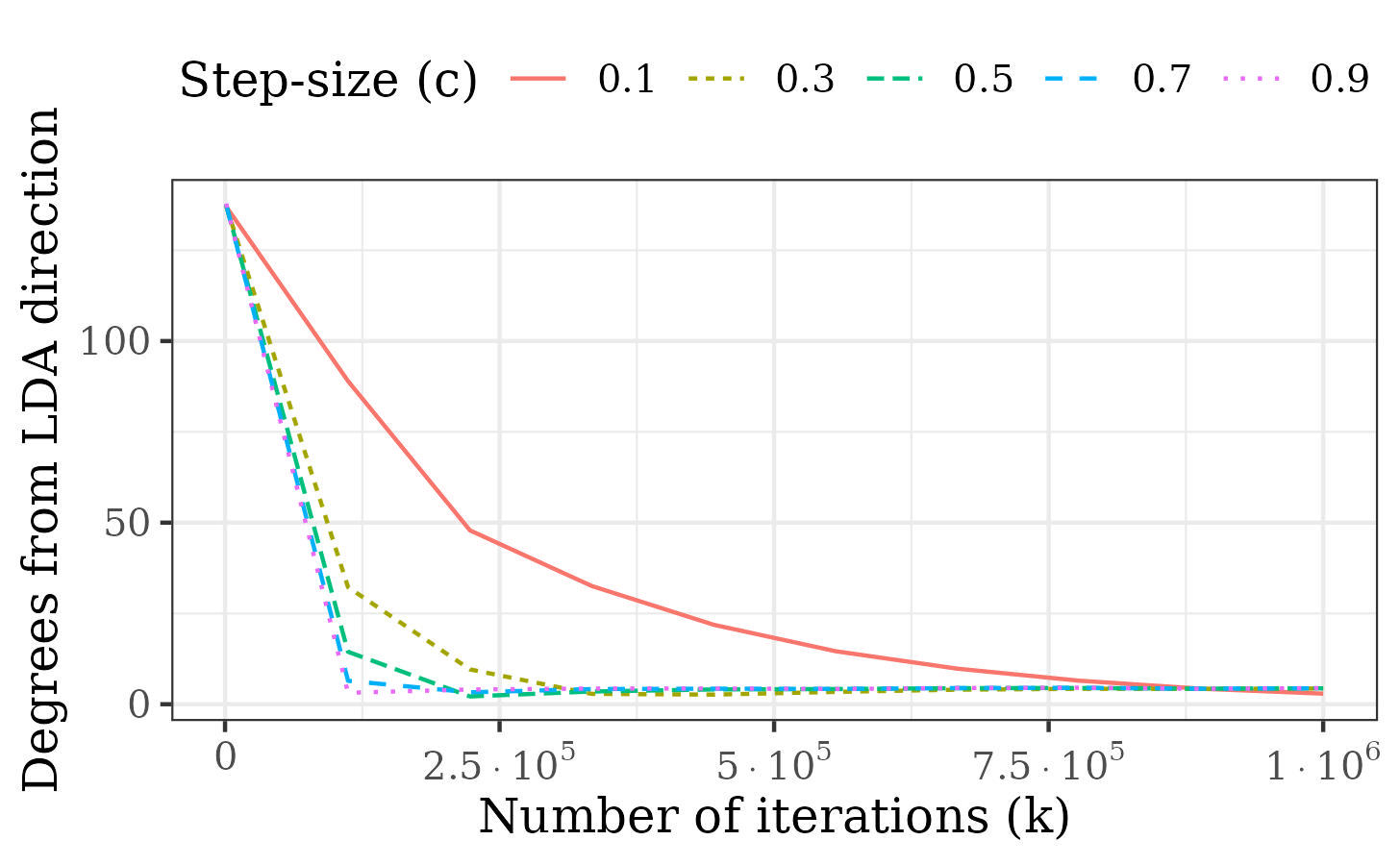}
		\caption{\begin{footnotesize}\label{fig:occupancy_angle}\emph{Mean angle (degrees) between LDA and rkLDA classifiers. }\end{footnotesize} }
	\end{subfigure}%
	\hspace{0.01\linewidth}
	\begin{subfigure}{.47\textwidth}
		\centering
		\includegraphics[width=\linewidth]{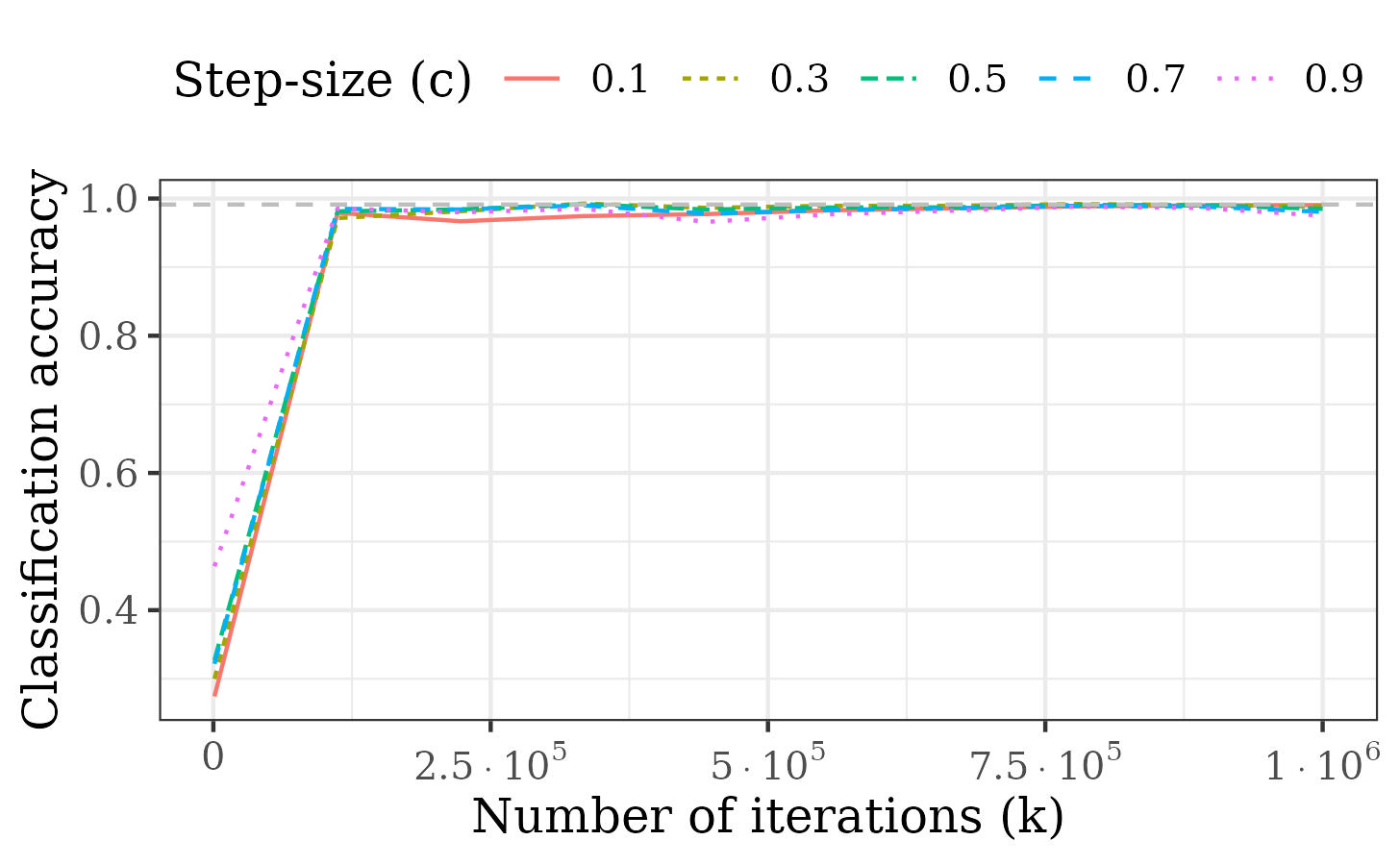}
		\caption{\begin{footnotesize}\label{fig:occupancy_accuracy}\emph{Mean rkLDA accuracy.  Gray line depicts gmLDA accuracy ($0.991$).}\end{footnotesize} }
	\end{subfigure}
	\caption{\begin{footnotesize}\emph{Occupancy detection results over 20 replicates as a function of iterations.}\end{footnotesize}}
	\label{fig:occupancydetection}
\end{figure}

Figure \ref{fig:occupancy_angle} depicts mean angle (in degrees) between gmLDA and rkLDA vectors from 20 replicates as a function of iterations $k$.  Colored lines depict results from different step-sizes $c$.  Results with all step-sizes appear to approach the gmLDA classifier and larger step-sizes converge more quickly to a solution that is close to the gmLDA classifier.  
Figure \ref{fig:occupancy_accuracy} depicts mean overall classification accuracy of the rkLDA solution from 20 replicates with the optimal intercept as a function of iterations $k$.  Colored lines depict results from different step-sizes $c$ while the dashed gray line depicts gmLDA accuracy.  Again, accuracy with all step-sizes $c$ approach gmLDA accuracy even before the angle from the gmLDA classifier is small. 

\subsection{Larger $p$ experiments illustrating accuracy and timing}\label{sec:exp2}

We present data and results from larger $p$ experiments illustrating accuracy and timing with comparisons to matrix sketching (Section \ref{sec:relatedworks}) and CG.

\subsubsection{Modified National Institute of Standards and Technology (MNIST) Dataset}

We employ the MNIST dataset from \cite{deng2012mnist} containing $70{,}000$ images with $p=784$ pixels ($28 \times 28$) each.  To identify similar digits for a more representative comparison, we perform PCA on all the MNIST digits and plot their representations in the first and second principal components.  Observing that the digits $6$ and $8$ exhibit substantial overlapping points in the first and second principal components, we subset the data for these two digits, resulting in $n=13{,}701$ images.  The training and testing datasets have scaled condition numbers $1.1 \times 10^{11}$ and $1.0 \times 10^{11}$, respectively.  This is largely due to the training and test datasets having squared Frobenius norm equal to $3.5 \times 10^{10}$ and $8.8 \times 10^{9}$, respectively.  In this data, $78$ percent of the entries are zero.

\begin{table}[ht]
	\setlength{\tabcolsep}{2pt}
	\centering
	\scalebox{0.78}{
		\begin{tabular}{lrrrrrrrrrr}  
			\hline 
			Method & \multicolumn{2}{c}{k=500} & \multicolumn{2}{c}{k=1{,}000} & \multicolumn{2}{c}{k=1{,}500} & \multicolumn{2}{c}{k=2{,}000} & \multicolumn{2}{c}{k=2{,}500} \\   
			\hline 
			Full data gmLDA & 0.9836 & & -- & & -- & & -- && -- & \\ 
			Full data CG & 0.9839  & & -- & & -- & & -- && -- & \\ 
			\hline 
			\multicolumn{11}{l}{\emph{Uniform weights}} \\
			rkLDA c=0.1 & 0.9787 & \emph{(0.0021)} & 0.9806 & \emph{(0.0015)} & 0.9813 & \emph{(0.0011)} & 0.9819 & \emph{(0.0014)} & 0.9827 & \emph{(0.0014)} \\   
			rkLDA c=0.3 & \textbf{0.9800} & \emph{(0.0025)} & \textbf{0.9816} & \emph{(0.0019)} & \textbf{0.9826} & \emph{(0.0018)} & \textbf{0.9827} & \emph{(0.0017)} & \textbf{0.9829} & \emph{(0.0015)} \\   
			rkLDA c=0.5 & 0.9790 & \emph{(0.0031)} & 0.9811 & \emph{(0.0023)} & 0.9810 & \emph{(0.0021)} & 0.9820 & \emph{(0.0024)} & 0.9817 & \emph{(0.0019)} \\   
			rkLDA c=0.7 & 0.9771 & \emph{(0.0035)} & 0.9793 & \emph{(0.0033)} & 0.9799 & \emph{(0.0027)} & 0.9793 & \emph{(0.0028)} & 0.9799 & \emph{(0.0025)} \\   
			rkLDA c=0.9 & 0.9744 & \emph{(0.0045)} & 0.9770 & \emph{(0.0034)} & 0.9772 & \emph{(0.0038)} & 0.9766 & \emph{(0.0052)} & 0.9770 & \emph{(0.0035)} \\   
			MSLDA & 0.6682 & \emph{(0.0438)} & 0.9318 & \emph{(0.0069)} & 0.9558 & \emph{(0.0043)} & 0.9665 & \emph{(0.0038)} & 0.9708 & \emph{(0.0033)} \\ 
			\hline 
			\multicolumn{11}{l}{\emph{Row-based weights}} \\
			rkLDA c=0.1 & 0.9777 & \emph{(0.0018)} & 0.9803 & \emph{(0.0016)} & 0.9812 & \emph{(0.0022)} & \textbf{0.9831} & \emph{(0.0012)} & 0.9826 & \emph{(0.0006)} \\   
			rkLDA c=0.3 & \textbf{0.9809} & \emph{(0.0020)} & \textbf{0.9821} & \emph{(0.0011)} & \textbf{0.9831} & \emph{(0.0014)} & 0.9821 & \emph{(0.0008)} & \textbf{0.9837} & \emph{(0.0012)} \\   
			rkLDA c=0.5 & 0.9779 & \emph{(0.0024)} & 0.9788 & \emph{(0.0048)} & 0.9815 & \emph{(0.0017)} & 0.9819 & \emph{(0.0012)} & 0.9828 & \emph{(0.0013)} \\   
			rkLDA c=0.7 & 0.9756 & \emph{(0.0028)} & 0.9796 & \emph{(0.0028)} & 0.9815 & \emph{(0.0003)} & 0.9799 & \emph{(0.0033)} & 0.9802 & \emph{(0.0039)} \\   
			rkLDA c=0.9 & 0.9737 & \emph{(0.0021)} & 0.9792 & \emph{(0.0032)} & 0.9765 & \emph{(0.0024)} & 0.9752 & \emph{(0.0034)} & 0.9776 & \emph{(0.0021)} \\   
			MSLDA & 0.6735 & \emph{(0.0461)} & 0.9315 & \emph{(0.0071)} & 0.9561 & \emph{(0.0045)} & 0.9655 & \emph{(0.0037)} & 0.9704 & \emph{(0.0034)} \\ 
			\hline
		\end{tabular}
	}
	\vspace{-.6\baselineskip}
	\caption{\begin{footnotesize}\emph{MNIST Accuracy Results -- Mean classification accuracy (and standard deviations) over varying values of $k$ (RK iterates for rkLDA and sketched dimension for MSLDA) and $100$ replicates.  Full data LDA and conjugate gradient (CG) results do not utilize $k$ and produce same results in all replicates. Most accurate mean results bolded in each of the uniform and row-based weights blocks.}\end{footnotesize}\label{table:mnistacc2}}
\end{table}

\begin{table}[ht]
	\setlength{\tabcolsep}{2pt}
	\centering
	\scalebox{0.78}{
		\begin{tabular}{lrrrrrrrrrr}  
			\hline 
			Method & \multicolumn{2}{c}{k=500} & \multicolumn{2}{c}{k=1{,}000} & \multicolumn{2}{c}{k=1{,}500} & \multicolumn{2}{c}{k=2{,}000} & \multicolumn{2}{c}{k=2{,}500} \\   
			\hline 
			Full data gmLDA  & 0.9388 & \emph{(0.0742)}  & -- & & -- & & -- && -- & \\ 
			Full data CG & 1.8179 & \emph{(0.0583)}  & -- & & -- & & -- && -- & \\ 
			\hline 
			\multicolumn{11}{l}{\emph{Uniform weights}} \\			
			rkLDA c=0.1 & 0.0985 & \emph{(0.0009)} & 0.1077 & \emph{(0.0036)} & \textbf{0.1131} & \emph{(0.0014)} & \textbf{0.1246} & \emph{(0.0074)} & 0.1296 & \emph{(0.0102)} \\   
			rkLDA c=0.3 & 0.0986 & \emph{(0.0008)} & 0.1078 & \emph{(0.0043)} & 0.1132 & \emph{(0.0010)} & 0.1249 & \emph{(0.0077)} & 0.1296 & \emph{(0.0104)} \\   
			rkLDA c=0.5 & 0.0986 & \emph{(0.0009)} & 0.1074 & \emph{(0.0026)} & \textbf{0.1131} & \emph{(0.0009)} & 0.1249 & \emph{(0.0076)} & 0.1298 & \emph{(0.0114)} \\   
			rkLDA c=0.7 & 0.0986 & \emph{(0.0021)} & \textbf{0.1073} & \emph{(0.0023)} & \textbf{0.1131} & \emph{(0.0009)} & 0.1251 & \emph{(0.0079)} & 0.1297 & \emph{(0.0107)} \\   
			rkLDA c=0.9 & 0.0984 & \emph{(0.0007)} & 0.1075 & \emph{(0.0022)} & \textbf{0.1131} & \emph{(0.0012)} & 0.1249 & \emph{(0.0076)} & \textbf{0.1294} & \emph{(0.0102)} \\   
			MSLDA & \textbf{0.0601} & \emph{(0.0033)} & 0.1228 & \emph{(0.0055)} & 0.1697 & \emph{(0.0029)} & 0.2189 & \emph{(0.0025)} & 0.2583 & \emph{(0.0027)} \\ 
			\hline 
			\multicolumn{11}{l}{\emph{Row-based weights}} \\
			rkLDA c=0.1 & 0.1034 & \emph{(0.0011)} & 0.1112 & \emph{(0.0005)} & 0.1223 & \emph{(0.0011)} & 0.1356 & \emph{(0.0015)} & 0.1424 & \emph{(0.0015)} \\   
			rkLDA c=0.3 & 0.1039 & \emph{(0.0008)} & 0.1112 & \emph{(0.0012)} & 0.1217 & \emph{(0.0011)} & 0.1352 & \emph{(0.0012)} & 0.1423 & \emph{(0.0010)} \\   
			rkLDA c=0.5 & 0.1086 & \emph{(0.0096)} & 0.1132 & \emph{(0.0041)} & \textbf{0.1212} & \emph{(0.0009)} & 0.1352 & \emph{(0.0019)} & \textbf{0.1417} & \emph{(0.0007)} \\   
			rkLDA c=0.7 & 0.1041 & \emph{(0.0008)} & \textbf{0.1111} & \emph{(0.0009)} & 0.1217 & \emph{(0.0011)} & \textbf{0.1351} & \emph{(0.0009)} & 0.1421 & \emph{(0.0009)} \\   
			rkLDA c=0.9 & 0.1044 & \emph{(0.0004)} & \textbf{0.1111} & \emph{(0.0005)} & 0.1215 & \emph{(0.0005)} & 0.1355 & \emph{(0.0003)} & 0.1430 & \emph{(0.0012)} \\   
			MSLDA & \textbf{0.0604} & \emph{(0.0056)} & 0.1237 & \emph{(0.0067)} & 0.1719 & \emph{(0.0076)} & 0.2201 & \emph{(0.0095)} & 0.2606 & \emph{(0.0110)} \\ 
			\hline
		\end{tabular}
	}
	\vspace{-.6\baselineskip}
	\caption{\begin{footnotesize}\emph{MNIST Timing Results -- Mean classification time in seconds (and standard deviations) over varying values of $k$ (RK iterates for rkLDA and sketched dimension for MSLDA) and $100$ replicates.  Full data LDA and conjugate gradient (CG) over $5$ replicates do not utilize $k$.  Fastest mean results bolded in each of the uniform and row-based weights blocks.}\end{footnotesize}\label{table:mnisttime}}
\end{table}

Tables \ref{table:mnistacc2} and \ref{table:mnisttime} depict mean classification accuracy and time results (with standard deviations in italics and parentheses) from full data gmLDA, full data CG, rkLDA, and MSLDA.  The gmLDA and CG solutions are non-random and therefore unchanged over replicates.  The rkLDA and MSLDA results are over 100 replicates for each value of $k$ (RK iterates for rkLDA and sketched dimension for MSLDA).   The middle section employs uniform sampling weights while the bottom employs row-based sampling weights, where the $i^{th}$ row was sampled with probability $p_{i} = \frac{\|\vx_{i}\|^{2}_{2}}{\|\mx\|^{2}_{F}}$.

Accuracy results show that rkLDA can obtain accuracy comparable to gmLDA even with a small number of iterations ($500 \le k \le2{,}500$) and even when the step-size is $0.9$, or close to $1$.  Timing results show that rkLDA achieves these results about 10 times faster than gmLDA on this dataset.  MSLDA is faster than rkLDA at smaller $k$ but suffers worse accuracy.  As $k$ increases, sketching accuracy drastically improves but also requires more time.  On this dataset, sketching is generally slower than rkLDA with all step-sizes, and less accurate at even the largest sketching dimension and for both uniform and row-based sampling weights.  Finally, CG has similar accuracy as gmLDA on this dataset but was substantially slower than rkLDA and matrix sketching at all values of $k$.

\begin{figure}[h]
	\centering
	\includegraphics[width=\linewidth]{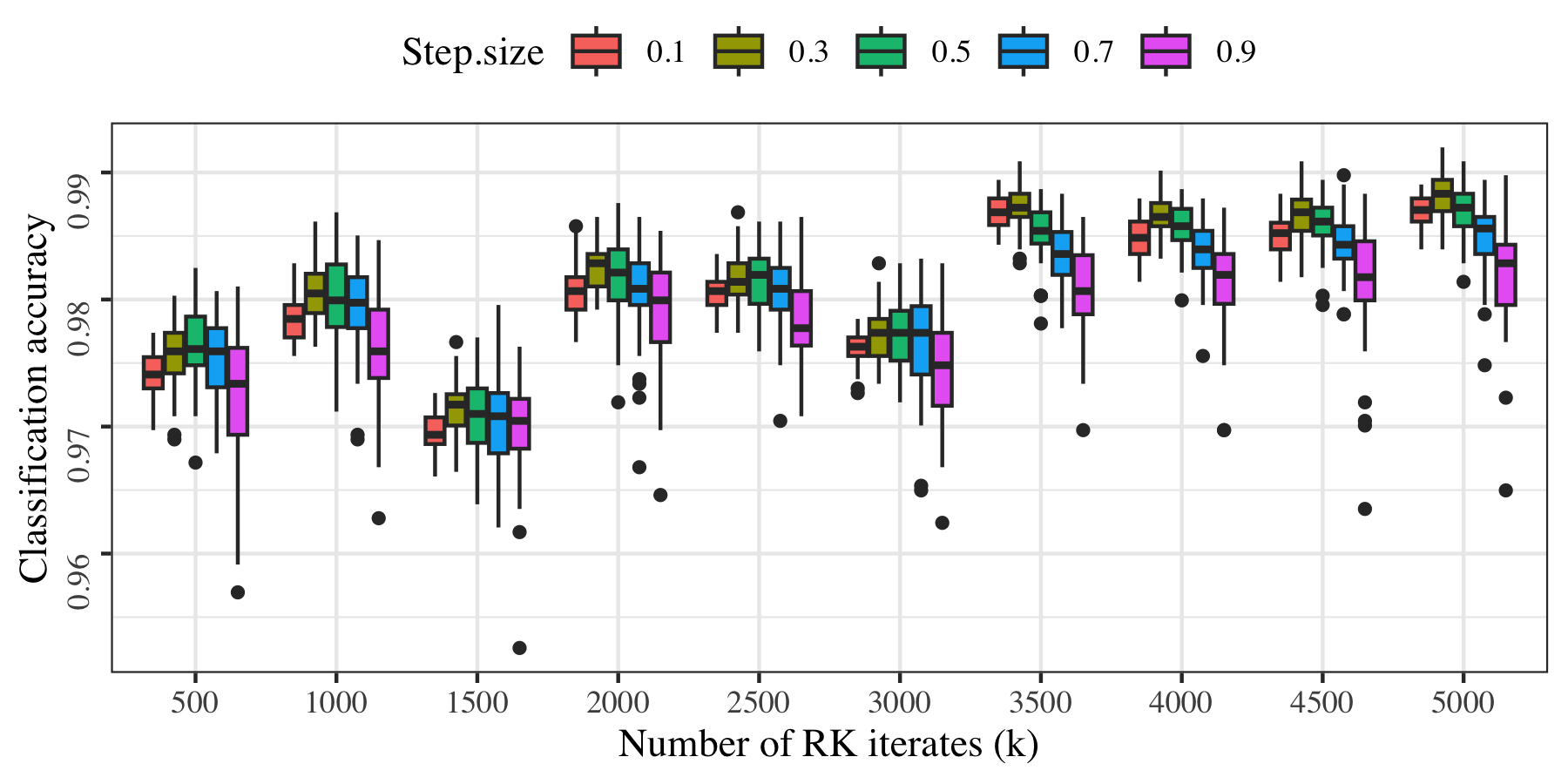}
	\vspace{-1.8\baselineskip}
	\caption{\begin{footnotesize}\emph{MNIST Classification Accuracy -- Boxplots of the classification accuracy for varying RK iterates $k$ and step-sizes $c$ over $50$ replicates.  Each set of $5$ boxplots over each value of $k$ indicates results for rkLDA with $c=0.1, 0.3, 0.5, 0.7$ and $0.9$ (from left to right).}\end{footnotesize}\label{fig:mnistconv}}
\end{figure}

Figure \ref{fig:mnistconv} depicts boxplots of the classification accuracy 
at the $k^{th}$ iterate 
for varying values of $k$ and step-sizes $c$ over $50$ replicates and row-based weights.  Table \ref{table:mnistacc2} shows that this was a very good problem for full data LDA and Figure \ref{fig:mnistconv} shows visually how similarly high accuracy results can be obtained with a relatively small number of RK iterates $k$.  Note that the values are not identical to those in Table \ref{table:mnistacc2} since the results are from two different sets of experiments. 

\subsubsection{CIFAR-10 Dataset}

We employ the CIFAR-10 dataset from \cite{krizhevsky2009learning} containing $50{,}000$ training and $10{,}000$ test images with $p=3{,}072$ pixels ($1{,}024$ entries for each of the three RGB channels).  To identify similar classes for a more representative comparison, we perform PCA on all the CIFAR-10 training images and plot their representations in the first and second principal components.  Observing that the groups containing ships and trucks exhibit substantial overlapping points in the first and second principal components, we subset the data for these two groups, resulting in $10{,}000$ training images and $2{,}000$ test images, equally split between the two classes.  The training and testing datasets have 
scaled condition numbers $3.3 \times 10^{11}$ and $5.3 \times 10^{10}$, respectively.  This is largely due the training and test datasets having squared Frobenius norm equal to $6.5 \times 10^{11}$ and $1.3 \times 10^{11}$, respectively.  Compared to MNIST, this dataset is dense; less than 0.1 percent of the entries are zero.  Since $k < p$ at all choices of $k$, we employ \verb|lsqr|, rather than the default solver, in the Scikit-learn  \verb|LinearDiscriminantAnalysis| module to perform gmLDA after matrix sketching.

\begin{table}[ht]
	\setlength{\tabcolsep}{2pt}
	\centering
	\scalebox{0.78}{
		\begin{tabular}{lrrrrrrrrrr}  
			\hline 
			Method & \multicolumn{2}{c}{k=500} & \multicolumn{2}{c}{k=1{,}000} & \multicolumn{2}{c}{k=1{,}500} & \multicolumn{2}{c}{k=2{,}000} & \multicolumn{2}{c}{k=2{,}500} \\   
			\hline 
			Full data gmLDA & 0.7320  & & -- & & -- & & -- && -- & \\ 
			Full data CG & 0.6670  & & -- & & -- & & -- && -- & \\ 
			\hline 
			\multicolumn{11}{l}{\emph{Uniform weights}} \\
			rkLDA c=0.1 & 0.6618 & \emph{(0.0623)} & 0.7021 & \emph{(0.0373)} & \textbf{0.7125} & \emph{(0.0327)} & \textbf{0.7264} & \emph{(0.0276)} & \textbf{0.7326} & \emph{(0.0223)} \\   
			rkLDA c=0.3 & \textbf{0.6761} & \emph{(0.0508)} & \textbf{0.7049} & \emph{(0.0481)} & 0.7093 & \emph{(0.0529)} & 0.7263 & \emph{(0.0416)} & 0.7241 & \emph{(0.0432)} \\   
			rkLDA c=0.5 & 0.6745 & \emph{(0.0626)} & 0.6900 & \emph{(0.0615)} & 0.6975 & \emph{(0.0625)} & 0.6992 & \emph{(0.0590)} & 0.6969 & \emph{(0.0710)} \\   
			rkLDA c=0.7 & 0.6461 & \emph{(0.0700)} & 0.6726 & \emph{(0.0719)} & 0.6720 & \emph{(0.0755)} & 0.6720 & \emph{(0.0744)} & 0.6901 & \emph{(0.0721)} \\   
			rkLDA c=0.9 & 0.6467 & \emph{(0.0722)} & 0.6547 & \emph{(0.0701)} & 0.6649 & \emph{(0.0689)} & 0.6628 & \emph{(0.0711)} & 0.6778 & \emph{(0.0677)} \\   
			MSLDA & 0.4980 & \emph{(0.0261)} & 0.4980 & \emph{(0.0205)} & 0.4985 & \emph{(0.0188)} & 0.5007 & \emph{(0.0167)} & 0.5011 & \emph{(0.0162)} \\ 
			\hline 
			\multicolumn{11}{l}{\emph{Row-based weights}} \\
			rkLDA c=0.1 & 0.6690 & \emph{(0.0510)} & \textbf{0.7030} & \emph{(0.0361)} & \textbf{0.7238} & \emph{(0.0262)} & \textbf{0.7261} & \emph{(0.0240)} & \textbf{0.7311} & \emph{(0.0253)} \\   
			rkLDA c=0.3 & \textbf{0.6797} & \emph{(0.0577)} & 0.7015 & \emph{(0.0517)} & 0.7081 & \emph{(0.0467)} & 0.7228 & \emph{(0.0442)} & 0.7223 & \emph{(0.0462)} \\   
			rkLDA c=0.5 & 0.6783 & \emph{(0.0634)} & 0.6828 & \emph{(0.0616)} & 0.6964 & \emph{(0.0580)} & 0.7127 & \emph{(0.0516)} & 0.6973 & \emph{(0.0628)} \\   
			rkLDA c=0.7 & 0.6466 & \emph{(0.0665)} & 0.6607 & \emph{(0.0749)} & 0.6755 & \emph{(0.0687)} & 0.6772 & \emph{(0.0742)} & 0.6753 & \emph{(0.0649)} \\   
			rkLDA c=0.9 & 0.6321 & \emph{(0.0779)} & 0.6653 & \emph{(0.0727)} & 0.6610 & \emph{(0.0712)} & 0.6630 & \emph{(0.0687)} & 0.6607 & \emph{(0.0773)} \\   
			MSLDA & 0.5047 & \emph{(0.0277)} & 0.4998 & \emph{(0.0211)} & 0.4980 & \emph{(0.0155)} & 0.5009 & \emph{(0.0158)} & 0.4974 & \emph{(0.0184)} \\ 
			\hline
		\end{tabular}
	}
	\vspace{-.6\baselineskip}
	\caption{\begin{footnotesize}\emph{CIFAR Accuracy Results -- Mean classification accuracy (and standard deviations) over varying values of $k$ (RK iterates for rkLDA and sketched dimension for MSLDA) and $100$ replicates.  Full data LDA and conjugate gradient (CG) results do not utilize $k$ and produce same results in all replicates. Most accurate mean results bolded in each of the uniform and row-based weights blocks.}\end{footnotesize}\label{table:cifaracc}}
\end{table}

\begin{table}[ht]
	\setlength{\tabcolsep}{2pt}
	\centering
	\scalebox{0.77}{
		\begin{tabular}{lrrrrrrrrrr}  
			\hline 
			Method & \multicolumn{2}{c}{k=500} & \multicolumn{2}{c}{k=1{,}000} & \multicolumn{2}{c}{k=1{,}500} & \multicolumn{2}{c}{k=2{,}000} & \multicolumn{2}{c}{k=2{,}500} \\   
			\hline 
			Full data gmLDA & 11.5235 & \emph{(0.1246)}  & -- & & -- & & -- && -- & \\ 
			Full data CG & 116.2896 & \emph{(2.8152)}  & -- & & -- & & -- && -- & \\ 
			\hline 
			\multicolumn{11}{l}{\emph{Uniform weights}} \\
			rkLDA c=0.1 & \textbf{0.8090} & \emph{(0.0215)} & 0.8305 & \emph{(0.0065)} & \textbf{0.8489} & \emph{(0.0142)} & 0.8585 & \emph{(0.0146)} & 0.8670 & \emph{(0.0151)} \\   
			rkLDA c=0.3 & 0.8093 & \emph{(0.0222)} & 0.8310 & \emph{(0.0084)} & 0.8505 & \emph{(0.0234)} & \textbf{0.8561} & \emph{(0.0069)} & 0.8699 & \emph{(0.0196)} \\   
			rkLDA c=0.5 & 0.8091 & \emph{(0.0213)} & 0.8307 & \emph{(0.0068)} & 0.8573 & \emph{(0.0561)} & 0.8604 & \emph{(0.0257)} & 0.8723 & \emph{(0.0457)} \\   
			rkLDA c=0.7 & 0.8144 & \emph{(0.0514)} & \textbf{0.8300} & \emph{(0.0041)} & 0.8533 & \emph{(0.0580)} & 0.8566 & \emph{(0.0099)} & \textbf{0.8659} & \emph{(0.0112)} \\   
			rkLDA c=0.9 & \textbf{0.8090} & \emph{(0.0198)} & 0.8303 & \emph{(0.0084)} & 0.8541 & \emph{(0.0531)} & 0.8611 & \emph{(0.0300)} & 0.8697 & \emph{(0.0226)} \\   
			MSLDA & 3.8442 & \emph{(0.0825)} & 3.9919 & \emph{(0.0866)} & 4.1563 & \emph{(0.0969)} & 4.3069 & \emph{(0.1007)} & 4.4227 & \emph{(0.1022)} \\
			\hline 
			\multicolumn{11}{l}{\emph{Row-based weights}} \\
			rkLDA c=0.1 & 0.8352 & \emph{(0.0235)} & \textbf{0.8467} & \emph{(0.0062)} & 0.8652 & \emph{(0.0240)} & \textbf{0.8934} & \emph{(0.0111)} & 0.9144 & \emph{(0.0171)} \\   
			rkLDA c=0.3 & \textbf{0.8311} & \emph{(0.0050)} & 0.8470 & \emph{(0.0140)} & 0.8686 & \emph{(0.0564)} & 0.8965 & \emph{(0.0254)} & \textbf{0.9126} & \emph{(0.0079)} \\   
			rkLDA c=0.5 & 0.8383 & \emph{(0.0486)} & 0.8496 & \emph{(0.0192)} & 0.8618 & \emph{(0.0068)} & 0.8979 & \emph{(0.0494)} & 0.9129 & \emph{(0.0141)} \\   
			rkLDA c=0.7 & 0.8371 & \emph{(0.0493)} & 0.8483 & \emph{(0.0182)} & 0.8634 & \emph{(0.0166)} & 0.8941 & \emph{(0.0126)} & 0.9130 & \emph{(0.0129)} \\   
			rkLDA c=0.9 & 0.8372 & \emph{(0.0483)} & \textbf{0.8467} & \emph{(0.0065)} & \textbf{0.8615} & \emph{(0.0092)} & 0.8978 & \emph{(0.0342)} & 0.9137 & \emph{(0.0107)} \\   
			MSLDA & 3.8925 & \emph{(0.0564)} & 4.0425 & \emph{(0.0501)} & 4.2091 & \emph{(0.0638)} & 4.3733 & \emph{(0.1116)} & 4.4724 & \emph{(0.0678)} \\
			\hline
		\end{tabular}
	}
	\vspace{-.6\baselineskip}
	\caption{\begin{footnotesize}\emph{CIFAR Timing Results -- Mean classification time in seconds (and standard deviations) over varying values of $k$ (RK iterates for rkLDA and sketched dimension for MSLDA) and $100$ replicates.  Full data LDA and conjugate gradient (CG) over $5$ replicates do not utilize $k$.  Fastest mean results bolded in each of the uniform and row-based weights blocks.}\end{footnotesize}\label{table:cifartime}}
\end{table}

Tables \ref{table:cifaracc} and \ref{table:cifartime} depict mean classification accuracy and time results (with standard deviations in italics and parentheses) from full data gmLDA, full data CG, rkLDA, and MSLDA.  Accuracy results show that rkLDA can obtain accuracy comparable to gmLDA even with a small number of iterations ($2{,}000\le k\le 2{,}500$).  In some cases, rkLDA can even outperform gmLDA, as the mean accuracy in the $c=0.1$ case exceeds gmLDA for $k \ge 2{,}500$.   Timing results show that rkLDA is about 10 times faster than gmLDA on this dataset.  MSLDA with the same $k$ as in rkLDA does not perform well here since $k<p$ in all cases and neither random row sampling nor gmLDA are intended for underdetermined systems.  CG with default tolerance for stopping condition of $10^{-5}$ is less accurate than both gmLDA and mean rkLDA for some choices of $c$ at all $k$ on this dataset.  Decreasing the tolerance increases runtime but does not improve accuracy on this dataset; CG does not converge at a tolerance of $10^{-13}$.  CG is also substantially slower than all other methods at all levels of $k$, likely because this dataset contains very little sparsity.

\begin{figure}[h]
	\centering
	\includegraphics[width=\linewidth]{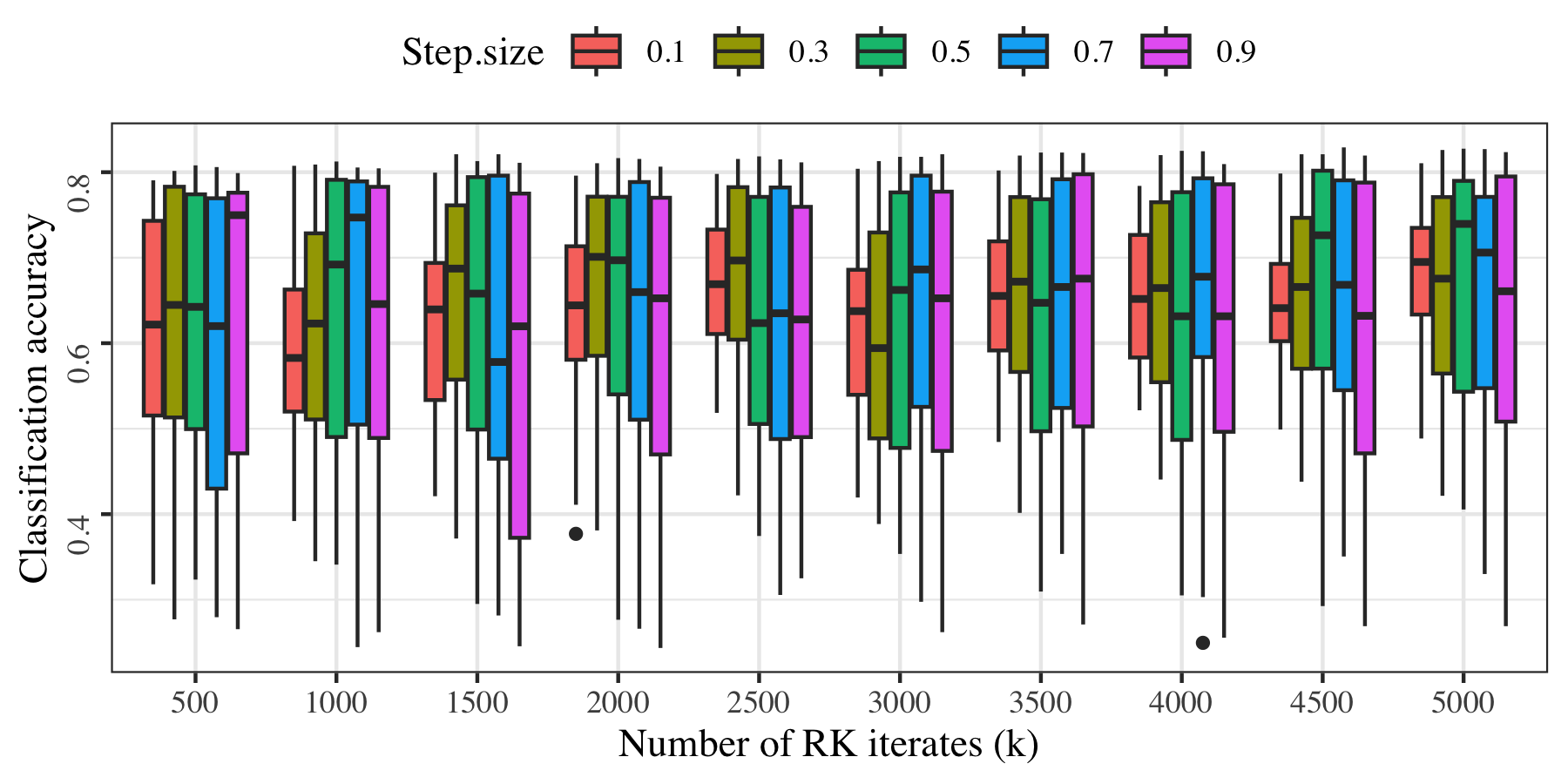}
	\vspace{-1.8\baselineskip}
	\caption{\begin{footnotesize}\emph{CIFAR-10 Classification Accuracy -- Boxplots of the classification accuracy for varying RK iterates $k$ and step-sizes $c$ over $50$ replicates.  Each set of $5$ boxplots over each value of $k$ indicates results for rkLDA with $c=0.1, 0.3, 0.5, 0.7$ and $0.9$ (from left to right).}\end{footnotesize}\label{fig:cifarconv}}
\end{figure}

Figure \ref{fig:cifarconv} depicts boxplots of the classification accuracy at the $k^{th}$ iterate for varying values of $k$ and step-sizes $c$ over $50$ replicates and row-based weights.  Table \ref{table:cifaracc} shows that this was a difficult problem for full data LDA and Figure \ref{fig:cifarconv} shows visually how similar accuracy results can be obtained with a relatively small number of RK iterates $k$.  Note that the values are not identical to those in Table \ref{table:cifaracc} since the results are from two different sets of experiments.

\section{Discussion}
\label{sec:discussion}

In this work, we present rkLDA, an iterative randomized approach to binary-class Gaussian model LDA for very large data.  The keys to our method and analysis lie in harnessing the least squares formulation of binary-class Gaussian model LDA, employing the machinery of the stochastic gradient descent framework with importance sampling \cite{needell2016stochastic}, and combining these with previous work on the statistical analysis of sketched linear regression \cite{ChiIpsenRandLS, MMY15}.  Therefore, our analysis of the expected deviation of the LDA discriminant function on new data when employing the RK solution in lieu of the least squares one after a finite number of steps account for \emph{both} randomness from the data modeling assumptions and algorithmic randomness from the RK iterates.

Although the RK is a very potent tool for solving large linear systems of equations, it does have some limitations.  First, the RK convergence rate depends on the conditioning of $\mx$ \cite[Theorem 2]{strohmer2009randomized} so that RK iterates from ill-conditioned $\mx$ converge more slowly.  However, our numerical experiments demonstrate that a relatively small number of iterations may be sufficient to obtain classification accuracy that is very comparable to that of LDA -- even when the scaled condition number of the training data is relatively high.  Therefore, it is possible to obtain a sufficiently accurate $\vhbeta$ for classification purposes with a relatively small number of iterates.  Nonetheless, one possible direction of future work for enhancing the rkLDA convergence rate for ill-conditioned design matrices may be to employ an acceleration method such as randomized block Kaczmarz \cite{needell2014paved} -- which can work well if the rows of $\mx$ can be partitioned into well-conditioned blocks, or selectable set RK \cite{ssrk21} -- which leverages information about the iterates to obtain a suitable subset of equations to be sampled from in each iteration.  Additionally, the RK iterates are susceptible to noise.  In particular, the least squares formulation of LDA in \eqref{eqn:ls-LDA} assumes noise in the response since observations are class-conditionally Gaussian.  Therefore, the convergence horizon in Theorem \ref{thm:expecteddiffXbeta} depends on both the step-size $c$ and the size of the least squares residuals $r^\star$.

We comment briefly on the convergence of the RK to a horizon of the least squares solution.  Empirically, our numerical experiments show that in the context of classification, an approximate solution that is within a horizon of the least squares solution can perform quite well.  Moreover, it is possible to achieve such a sufficiently accurate solution with a relatively small number of iterates and without exceedingly small values for the step-size parameter $c$.  More sophisticated variations such as SGD with decreasing step-size or the randomized extended Kaczmarz \cite{zouzias2013randomized} are able to converge to the least squares solution.  To keep our contributions in applying an RK-type method to LDA clear, we focus on analyzing the RK without the need to include non-obvious decreasing step-size agendas or extension methods that require column-wise computations.   Nonetheless, it could be interesting future work to see whether or not there are additional gains in the classification context to be had from achieving a more accurate approximate solution.

While we consider the overdetermined $n>p$ case in this work, many applications such as image recognition and text analysis can involve $p>n$.  This ultrahigh dimension case is a well-known problem in Gaussian model LDA since $\mhsig\Inv$ in 
\eqref{eqn:discriminantfuncs-v3} does not exist in this case \cite{mai2013review} and there have been a number of works addressing this issue \cite{bickel2004some, pan2016ultrahigh, li2019multiclass, ni2016entropy}.  Therefore, another avenue of future work may be to adapt the analysis and optimal intercept in \eqref{eqn:optimalbetao} for ultrahigh dimensions.  Since \eqref{eqn:optimalbetao} involves computing $\mhsig \in \mathbb{R}^{p \times p}$, this may become computationally prohibitive for $p$ sufficiently large.  One possibility may be to employ randomized approaches to computing the sample covariance.  This would involve exploring how these approximations in the optimal intercept may affect classification accuracy. 

Finally, we briefly comment on how one can employ rkLDA when the data may be too large to store in working memory on a local machine.  Since Algorithm \ref{alg:lda-rk} only requires one row of the training data in each iteration, one may store and compute iterates on batches of sampled rows from the training data sequentially.  In this way, one bypasses the need to store the full data.  We note that unlike the leverage scores, the row norms and $\mhsig$ for the optimal intercept in \eqref{eqn:optimalbetao} can also be computed by viewing only batches of the data sequentially.  The idea of viewing batches of the data in sequence is similar to streaming discriminant analysis \cite{pang2005incremental, hayes2020lifelong, anagnostopoulos2012online}.  While these are viable alternatives to full data analysis in the streaming cases, a primary difference is that these lack theoretical convergence guarantees for how close the resulting classifier is to the original full data classifier.  Additionally, our numerical experiments demonstrate that rkLDA can be a very viable alternative to LDA in terms of both accuracy and timing when the entire training and test data fit in local memory.  Therefore, another promising line of future work would be to employ parallel RK algorithms \cite{kamath2015distributed, keinert2022randomized, liu2014asynchronous, wang2022prkp} to adapt rkLDA for distributed computing.

\begin{footnotesize}
\bibliographystyle{siamplain}
\bibliography{references}
\end{footnotesize}

\end{document}